\documentclass[a4paper]{ar-1col}
\usepackage[utf8]{inputenc}
\usepackage{url}
\usepackage[numbers]{natbib}
\usepackage{amsmath}
\usepackage{mathrsfs}
\usepackage{siunitx}
\usepackage{physics}
\usepackage{graphicx}
\usepackage{booktabs}

\newcommand{\bph}{\boldsymbol{\hat \phi} }

\newcommand{\epsilonb}{\boldsymbol{\epsilon}} 
\newcommand{\pre}{\text{(pre)}}

\makeatletter
\def\odd@review@icon@height{10pt}

\def\odd@review@icon#1{
	\let\annualreviewGin@setfile\Gin@setfile
	\let\Gin@setfile\oldGin@setfile
	\raisebox{-0.25\height}{\includegraphics[height=\odd@review@icon@height]{icons/#1.pdf}}
	\let\Gin@setfile\annualreviewGin@setfile
}

\def\so{\odd@review@icon{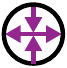}}
\def\st{\odd@review@icon{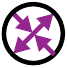}}
\def\p{\odd@review@icon{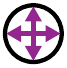}}
\def\tor{\odd@review@icon{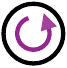}}
\def\eo{\odd@review@icon{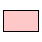}}
\def\et{\odd@review@icon{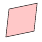}}
\def\es{\odd@review@icon{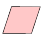}}
\def\di{\odd@review@icon{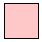}}
\def\rot{\odd@review@icon{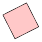}}
\def\dtV{\odd@review@icon{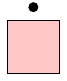}}
\def\dtR{\odd@review@icon{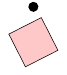}}
\def\dtSo{\odd@review@icon{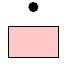}}
\def\dtSt{\odd@review@icon{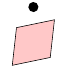}}
\def\dtS{\odd@review@icon{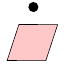}}

\makeatother

\usepackage{bm}
\usepackage{braket}
\usepackage{csquotes}
\def\vec#1{\bm{#1}}

\let\strong\textbf

\jname{Xxxx. Xxx. Xxx. Xxx.}
\jvol{AA}
\jyear{YYYY}
\doi{10.1146/((please add article doi))}

\usepackage{hyperref}
\usepackage{bookmark}

\hypersetup{%
  breaklinks=true,
  colorlinks=true,
  linkcolor=black,
  filecolor=black,
  urlcolor=black,
  citecolor=black,
}
\begin{document}

\markboth{Author et al.}{Odd viscosity and elasticity}

\title{Odd Viscosity and Odd Elasticity}

\author{Michel Fruchart,$^1$ Colin Scheibner,$^1$ and Vincenzo Vitelli$^{1,2}$
\affil{$^1$James Franck Institute and Department of Physics, The University of Chicago, Chicago, IL 60637, USA}
\affil{$^2$Kadanoff Center for Theoretical Physics, The University of Chicago, Chicago, IL 60637, USA; email: vitelli@uchicago.edu}
}
\begin{abstract}
Elasticity typically refers to a material’s ability to store energy, while viscosity refers to a material’s tendency to dissipate it. In this review, we discuss fluids and solids for which this is not the case. These materials display additional linear response coefficients known as odd viscosity and odd elasticity. We first introduce odd elasticity and odd viscosity from a continuum perspective, with an emphasis on their rich phenomenology, including transverse responses, modified dislocation dynamics, and topological waves. We then provide an overview  of systems that display odd viscosity and odd elasticity. These systems range from quantum fluids, to astrophysical gasses, to active and driven matter. Finally, we comment on microscopic mechanisms by which odd elasticity and odd viscosity arise.
\end{abstract}

\begin{keywords}
continuum mechanics, fluid mechanics, non-equilibrium, active matter, hydrodynamics 
\end{keywords}
\maketitle

\tableofcontents

\section{Introduction}

Continuum theories are phenomenological tools that allow us to understand and manipulate the world around us~\cite{Truesdell1960,Landau6,Landau7}. 
As it turns out, classical field theories such as elasticity and fluid mechanics have been built upon certain symmetries and conservation laws that describe disparate systems ranging from rivers to steel beams.
Examples include time-reversal invariance, mirror reflection symmetry, or conservation of energy and angular momentum.
Yet, not all systems are subject to these constraints: counterexamples range from biological tissues and astrophysical gases to quantum fluids. 
In this review, we analyze minimal extensions of fluid and solid mechanics that arise when the microscopic constituents of a medium do not follow the usual symmetries and conservation laws.
As a consequence, the continuum theories acquire new terms that take into account the collective effect of the broken symmetries. 
The advantage of this perspective is that we obtain universal effective theories that describe an array of systems that span many areas of research.
The drawback is that these universal theories, by nature, do not describing every detail of each system.

The key feature shared by all the systems in this review is that their evolution is governed by forces between adjacent parcels of a continuous medium. This physics is captured mathematically by the stress tensor
\begin{align}
\label{eq:stress_overview} 
    \sigma_{ij} = C_{ijk\ell} \partial_\ell u_k \, + \, \eta_{ijk\ell} \partial_\ell \dot u_k \, + \, \cdots 
\end{align}
that summarizes the surface forces between material elements. 
In terms of the stresses, the internal forces in the medium are given by $f_i= \partial_j \sigma_{ij}$. The elasticity tensor $C_{ijk\ell}$ is the proportionality coefficient between the stress tensor and the displacement gradient $\partial_\ell u_k$. The viscosity tensor $\eta_{ijk\ell}$ is the proportionality coefficient the stress and the velocity gradient $\partial_i \dot u_j$. Eq.~(\ref{eq:stress_overview}) gives a mechanical definition of elasticity and viscosity. In usual fluids and solids, respectively, these coefficients can be expressed as 
\begin{align}
    \label{viscosity_elasticity_potential}
    C_{i j k \ell} = \frac{\delta^2 F}{\delta (\partial_j u_i) \delta (\partial_\ell u_k ) } 
    \qquad
    \text{and}
    \qquad
    \eta_{i j k \ell} = \, \frac{T\delta^2 \dot{S}}{\delta ( \partial_j \dot u_i) \delta (\partial_\ell \dot u_k)}
    \qquad \text{(usually)}
\end{align}
where $F$ is the free energy of the elastic medium, $\dot S$ the rate of entropy production of the fluid, and $T$ is the temperature. 
Equation \eqref{viscosity_elasticity_potential} summarizes the usual meaning of elasticity and viscosity: elasticity usually describes the reversible storage of energy, while viscosity describes its irreversible dissipation. When 
Eq.~\eqref{viscosity_elasticity_potential} holds, the  elasticity and viscosity tensors obey the symmetry
\begin{equation}
\label{symmetric_C_eta_tensors}
C_{i j k \ell} = C_{k \ell i j}
\qquad
\text{and}
\qquad
\eta_{i j k \ell} = \eta_{k \ell i j} \qquad \text{(usually)}
\end{equation}
This review discusses situations in which Eq.~\eqref{symmetric_C_eta_tensors} is not valid, i.e., situations in which the elastic and viscous tensor are not symmetric:
\begin{equation}
\label{eq:oddelvis}
C_{i j k \ell} \neq C_{k \ell i j}
\qquad
\text{and}
\qquad
\eta_{i j k \ell} \neq \eta_{k \ell i j} \qquad \text{(more generally)}
\end{equation}
The antisymmetric parts of $C_{ijk\ell}$ and $\eta_{ijk\ell}$ in Eq.~(\ref{eq:oddelvis}) are henceforth referred to as ``odd" because they flip sign upon exchanging the pair of indices ${ij}$ with ${k\ell}$. 
As we will explain, odd elasticity is generally associated with microscopic non-conservative forces, while odd viscosity is usually associated with microscopic dynamics that do not obey time reversal symmetry. 
These coefficients arise is systems spanning scales, from quantum fluids to geophysical flows, and tabletop experiments with driven and active particles. 

\section{Odd viscosity}

\subsection{What is viscosity?}
When you stir a fluid, such as water or honey, the fluid resists the motion you are trying to impart. 
This is due to viscosity, which captures the resistance of the fluid to having inhomogeneities in its velocity field.
Honey is more viscous than water, so stirring honey requires more work than stirring water.
When velocity gradients are present, forces that tend to make these velocities equal appear between neighboring fluid parcels. 
Formally, this is described by the Navier-Stokes equation
\begin{equation}
    \label{ns}
    \rho D_t v_i = \partial_j \sigma_{i j} + f_i
\end{equation}
in which $D_t = \partial_t + v_k \partial_k$ is the convective derivative, $v(t, x)$ is the velocity field, $f_i$ are external body forces, and 
\begin{equation}
    \label{stress_strain_rate_relation}
    \sigma_{i j} = \sigma_{i j}^{\rm h} + \eta_{i j k \ell} \partial_\ell v_k
\end{equation}
is the stress tensor, which describes the forces between fluid parcels.
We have decomposed the stress tensor into so-called hydrostatic stresses $\sigma_{i j}^{\rm h}$ that are present even when the fluid is at rest (in simple fluids, $\sigma_{i j}^{\rm h} = - p \delta_{i j}$ where $p$ is the pressure), and viscous stresses $\sigma^{\rm vis}_{ij} = \eta_{i j k \ell} \partial_\ell v_k$, which are induced by velocity gradients\footnote{In general, the viscous stress tensor can be a non-linear function of the velocity gradients. Moreover, additional stresses can arise from the presence of gradients in other fields, such as temperature. These effects are ignored here for simplicity. }.
The coefficient of proportionality  $\eta_{i j k \ell}$ between velocity gradients and the stress is called the viscosity tensor.
The divergence of the viscous stress tensor is a force density $f_i^{\rm vis} = \partial_j \eta_{i j k \ell} \partial_\ell v_k$ that enters the Navier-Stokes equation.
In a isotropic, incompressible fluid at equilibrium, it takes the familiar form $f_i^{\rm vis} = \eta \, \Delta v_i$, where $\eta$ is the shear viscosity.

As one may expect from the example of water and honey,  
viscosity measures how much mechanical energy is converted into heat by the friction between layers of fluid. Indeed, the rate of loss of mechanical energy by viscous dissipation is~\cite{Landau6,Khain2022}
\begin{equation}
    \label{viscous_dissipation}
    \dot{w} = \sigma_{i j}^{\rm vis} \partial_j v_i = \eta_{i j k \ell} (\partial_j v_i) (\partial_\ell v_k) = \eta^{\rm S}_{i j k \ell} (\partial_j v_i) (\partial_\ell v_k).
\end{equation}
From this expression, we see that only the symmetric part $\eta^{\rm S}_{i j k \ell} = [\eta_{i j k \ell} + \eta_{k \ell i j}]/2$ of the viscosity tensor contributes to dissipation. The antisymmetric part 
\begin{equation}
    \eta^{\rm A}_{i j k \ell} = [\eta_{i j k \ell} - \eta_{k \ell i j}]/2
    \qquad
    \text{(odd viscosity tensor)}
\end{equation}
drops out. Therefore, the antisymmetric part of the viscosity tensor $\eta^{\rm A}_{i j k \ell}$ describes non-dissipative viscosities, that are called \emph{odd viscosities}~\cite{Avron1998}. 
Depending on the context, odd viscosity is known by other names, such as: Hall viscosities, gyroviscosities, Lorentz shear modulus, Senftleben-Beenakker effect.

\begin{marginnote}[]
\entry{Odd viscosity}{non-dissipative component of the viscosity tensor, contained in its antisymmetric part}
\end{marginnote}

\subsubsection{Two-dimensional fluids with odd viscosity}
\label{2d_fluids}
In two dimensions, the viscosity tensor $\eta_{ijk\ell}$ contains $2^4=16$ independent components. To keep track of them, it is useful to introduced a physically intuitive basis for stress and strain rate,  summarized in Table~\ref{table_ir_2d} and in the S.I. 
In this notation, the velocity gradient $\partial_i v_\ell$ and stress $\sigma_{ij}$ are represented by vectors $\dot e_\alpha$ and $\sigma_\alpha$, respectively, and the viscosity tensor $\eta_{ijk\ell}$ can be represented as a four-by-four matrix $\eta_{\alpha \beta}$ ($\alpha, \beta=0,\dots,3$).
In an isotropic fluid (i.e. one with no distinguished axis), the stress-velocity gradient takes the form~\cite{Han2021}
\begin{equation}
\label{general_stress_2d}
\begin{gathered}
\makeatletter
\let\annualreviewGin@setfile\Gin@setfile
\let\Gin@setfile\oldGin@setfile
\includegraphics[width=7cm]{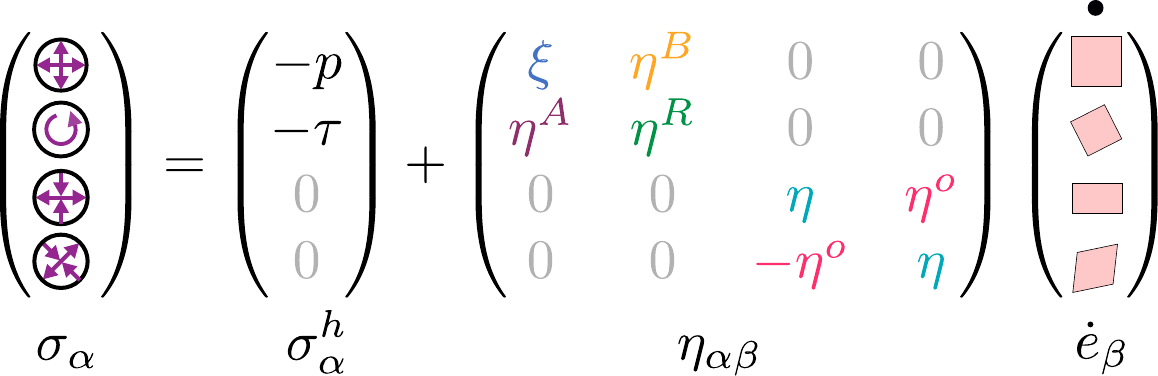}
\let\Gin@setfile\annualreviewGin@setfile
\makeatother
\end{gathered}
\end{equation}
Here, $p$ is the hydrostatic pressure and $\tau$ is an hydrostatic torque that can arise, for instance, when particles have transverse (i.e., non-central) forces. 
The matrix $\eta_{\alpha \beta}$ has six independent coefficients. The shear, bulk, and rotational viscosities $\eta$, $\zeta$, and $\eta^{\rm R}$ arise in usual fluids. 
The coefficients $\eta^{\rm A}$ and $\eta^{\rm B}$ couple compression and rotation, while the odd shear viscosity $\eta^{\rm o}$ couples the two independent shears. In standard tensor notation, Eq.~(\ref{general_stress_2d}) reads
\begin{equation}
\begin{split}
    \eta_{ijk\ell} = \;&\zeta \, \delta_{ij} \delta_{k\ell} -  \eta^\text{A} \epsilon_{ij} \delta_{k\ell} - \eta^\text{B} \delta_{ij} \epsilon_{k\ell}  + \eta^\text{R} \epsilon_{ij} \epsilon_{k\ell} \\
    + \, &\eta \, (\delta_{ik} \delta_{j\ell} + \delta_{i\ell} \delta_{jk} - \delta_{ij} \delta_{k\ell}) + \eta^\text{o} (\epsilon_{ik} \delta_{j\ell}+\epsilon_{j\ell} \delta_{ik})
\end{split}
\end{equation}
where $\delta_{ij}$ and $\epsilon_{ij}$ denote the Kronecker delta and Levi-Civita tensors (with $\epsilon_{x y} = 1$).
The symmetry $\eta_{ijk\ell}= \eta_{k\ell ij}$ is equivalent to the symmetry $\eta_{\alpha \beta} =\eta_{\beta \alpha}$ of the matrix in Eq.~\eqref{general_stress_2d}. 
Hence, having no odd viscosities ($\eta^{\rm A}_{ijk\ell} = 0$) amounts to having $\eta^{\rm A} = \eta^{\rm B}$ and $\eta^{\rm o}=0$. 
By coincidence, it also turn out that $\eta^{\rm A}$, $\eta^{\rm B}$, and $\eta^{\rm o}$ all violate mirror symmetry. Hence, in order to have odd viscosities in two dimensional isotropic fluids, it is necessary that some physical ingredient breaks mirror symmetry, for instance particles all rotating in the same direction or an external magnetic field.

Anisotropic 2D fluids can have more odd viscosities~\cite{Rao2020,Souslov2020}: this has been discussed in the context of nematic systems~\cite{Hess1986,Jarkova2001,Souslov2020} as well as in solid-state physics, where the viscosity tensor can be constrained by the crystallographic symmetries of the underlying lattice~\cite{Rao2020,Varnavides2020,Cook2021,Friedman2022,Huang2022}.
In the language of rheology, odd viscosity can be expressed in terms of the so-called normal stress difference (more precisely, the part that is odd in shear rate), see Ref.~\cite{Zhao2021} for details.  
Finally, we note that odd viscosities have also been studied in relativistic fluids~\cite{Nicolis2011,Jensen2012}.

\begin{table}[h]
    \caption{\label{table_ir_2d}
    Irreducible components of rank-two tensors in 2D.
    }
    {
    \centering
    \begin{tabular}{llll}
        \toprule
        deformation & deformation rate & stress & geometric meaning \\
        \midrule
        $e_0 = \di = \partial_x u_x + \partial_y u_y$ & $\dot{e}_0 = \dtV = \partial_x v_x + \partial_y v_y$ &  $\sigma_0 = \p  = [\sigma_{xx}+\sigma_{yy}]/2$ & isotropic area change \\
        $e_1 = \rot = \partial_x u_y - \partial_y u_x$ & $\dot{e}_1 = \dtR =  \partial_x v_y - \partial_y v_x$ & $\sigma_1=\tor=[\sigma_{yx}-\sigma_{xy}]/2$ & rotation \\
        $e_2 = \eo =  \partial_x u_x - \partial_y u_y$ & $\dot{e}_2 = \dtSo = \partial_x v_x - \partial_y v_y$ & $\sigma_2=\so=[\sigma_{xx}-\sigma_{yy}]/2$ & pure shear 1 \\
        $e_3 = \et = \partial_x u_y + \partial_y u_x$ & $\dot{e}_3 = \dtSt = \partial_x v_y + \partial_y v_x$ & $\sigma_3=\st=[\sigma_{xy}+\sigma_{yx}]/2$ & pure shear 2 \\
        \bottomrule
    \end{tabular}
    }
    \begin{tabnote}
    A pure shear (rate) correspond to a (rate of) change in shape without a change in volume or orientation.
    Shear 1 describes a horizontal elongation and vertical compression while shear 2 describes elongation along the \ang{45}  direction and compression along the \ang{-45} direction. 
    They are mathematically orthogonal.
    We note that $\dot{e}_0 = \nabla \cdot (\vec{v})$ and
    $\dot{e}_1 = \vec{\omega}$ is vorticity.
    The stresses are the conjugate forces to these deformations, and have similar interpretations.
    In particular, $\sigma_0$ includes pressure. The antisymmetric stress $\sigma_1$ is discussed in the Box~\emph{Antisymmetric Stress}.
    \end{tabnote}
\end{table}

\subsubsection{Three-dimensional fluids with odd viscosity} 
\label{3d_fluids}
In 3D, the situation is more complex. The tensor $\eta_{ijk\ell}$ has $3^4 =81$ independent components. It turns out that there are no odd viscosities compatible with spatial isotropy, implying that an odd viscous fluid in three dimensions must have (at the minimum) a preferred axis. A simple way to do that is to apply a magnetic field to the system, or to rotate it (this can indeed produce odd viscosity: see section \ref{sec_collisionless}). In this case the system still has cylindrical symmetry, and there are eight independent non-dissipative (i.e. odd) viscosities, six of which require parity to be broken (there are also eleven dissipative viscosities, three of which require parity to be broken); a full classification is given in Ref.~\cite{Khain2022}. This highlights that non-dissipative viscosities and parity-violating viscosities are distinct notions.
Like in 2D, point group symmetries of the crystal further constrain the viscosity tensor in solid-state systems~\cite{Rao2020,Varnavides2020,Robredo2021}.

\begin{textbox}[t]\section{Antisymmetric Stress}
The stress tensor of fluids and solids composed of particles with noncentral interactions can exhibit an antisymmetric part~\cite{Condiff1964}, the antisymmetric stress $\sigma^{\text{A}}_{ij} = \frac12 [\sigma_{ij} - \sigma_{ji}]$~\cite{Condiff1964,Han2021}. 
This effect is distinct from odd viscosity and odd elasticity, but often coexists with them. Mechanically, the antisymmetric stress corresponds to a torque density $\tau_i = \frac12 \epsilon_{ijk} \sigma^{\text{A}}_{jk}$, and it can arise, for instance, in fluids and solids made of spinning particles.
Noteably, it can exist even in the absence of velocity or displacement gradient, in which case it is then part of the so-called hydrostatic stress in fluids and of the pre-stress in solids. 
In addition, odd viscosity and elasticity can contribute to the antisymmetric stress (e.g. through $\eta^{\text{A}}$ in Eq.~(\eqref{general_stress_2d}) and $A$ in Eq.~\eqref{eq:modfam}).
An antisymmetric stress can also be generated by other mechanisms (distinct from odd elasticity or viscosities), such as passive Cosserat elasticity~\cite{Eringen2012} or rotational viscosities (such as $\eta^{\text{R}}$ in Eq.~\eqref{general_stress_2d}, see also \cite{deGroot1954}). 
Note that the phrase \emph{odd stress} (or \emph{Hall stress}) has been variously used in the literature to refer to both antisymmetric stress, or to any stress due to odd elasticity or viscosity.
\end{textbox}

\subsection{How does odd viscosity affect flows?}

\subsubsection{Incompressible 2D flows}

Let us first consider the case of an incompressible (low Mach number) two-dimensional flow, and consider only shear viscositites $\eta$ and $\eta^{\rm o}$ \footnote{In dilute systems, the rotational viscosities $\eta^{\rm A}$ and $\eta^{\rm B}$ are small and may be neglected~\cite{Han2021}.}. The Navier-Stokes equations for such a system are
\begin{align}
    \label{ns2d}
   \rho D_t \vec{v} =& - \nabla p + \eta \Delta \vec{v} + \eta^{\text{o}} \epsilonb \cdot \Delta \vec{v}
\end{align}
with $\nabla \cdot \vec{v} = 0$. 
In Eq.~\eqref{ns2d}, odd viscosity induces a viscous force perpendicular to the force one would generally expect from shear viscosity (the matrix $\epsilonb$ performs a clockwise rotation by \ang{90}). 
Because of the incompressibility of the fluid, 
the odd viscous term $\eta^o \epsilonb \cdot \Delta \vec{v}$ can be incorporated in an effective pressure~\cite{Ganeshan2017,Banerjee2017}
\begin{equation}
  \label{modified_pressure_2D}
  p' = p - \eta^{\text{o}} \omega
\end{equation}
where $\omega = \epsilon_{k\ell} \partial_k v_\ell$ is the vorticity of the fluid, and we used the identity $\epsilon \Delta \vec{v} = \nabla \omega$ for incompressible flows.
We can therefore map the Navier-Stokes equation with odd viscosity \eqref{ns2d} onto an effective system $\rho D_t \vec{v'} = - \nabla p' + \eta \Delta \vec{v'}$ without odd viscosity, with $\vec{v'} = \vec{v}$. 
This fact has an important consequences: when boundary conditions only involve the velocity of the fluid, the velocity field in a 2D fluid with and without $\eta^o$ are indistinguishable.

As an illustration, let us consider a two-dimensional Poiseuille flow, in which a fluid flows through a pipe under a constant pressure gradient.
In a normal fluid ($\eta^{\text{o}} = 0$), the velocity field is~\cite{Guyon2015}
\begin{equation}
  \vec{v} = \frac{G}{2 \eta} y (h-y) \vec{\hat{e}}_x
\end{equation}
where $G$ is the pressure gradient in the direction $x$ (with unit vector $\vec{\hat{e}}_x$), $h$ is the height of the channel, and $\eta$ is the (normal) shear viscosity. It satisfies the no-slip boundary conditions $\vec{v} = \vec{0}$ (which do not involve the stress) at the walls.
The vorticity is given by $\omega = \partial_x v_y - \partial_y v_x = - {G}/[{2 \eta}] \, (h - 2 y)$.
We now consider an fluid with odd viscosity in the same geometry.
Recall that the velocity field is not modified by the presence of odd viscosity (nor is the vorticity).
The pressure, however, is modified. Using Eq.~\eqref{modified_pressure_2D}, we find that it becomes
\begin{equation}
  p = p' + \eta^{\text{o}} \omega = p' - \frac{G \eta^{\text{o}}}{2 \eta} (h - 2 y)
\end{equation}
in which $p'$ is the pressure in the system without odd viscosity. 
The transverse pressure difference, measured between the bottom and the top of the channel, is then\footnote{In the 2D incompressible system, the difference between normal forces at the top and bottom wall  $\sigma_{yy}(h)-\sigma_{yy}(0)$ does not capture odd viscosity, because in this geometry $\sigma_{yy}(y) = - p - \eta^{\rm o} \partial_y v_x = - p'(y)$ is unchanged compared to the case with $\eta^{\rm o} = 0$. The changes from pressure and viscous shear stress that contribute to these normal forces compensate exactly.  }
\begin{equation}
  \Delta p = p(h) - p(0) = G \, h \,\frac{\eta^{\text{o}}}{\eta}
\end{equation}
as the pressure $p'(y)$ of the Poiseuille flow without odd viscosity is up-down symmetric.
A slightly more elaborate version of this method has been used to measure odd viscosities in three-dimensional polyatomic gases~\cite{Beenakker1970,Mccourt1990,Korving1966,Korving1967,Hulsman1970,Hulsman1970b}, see Fig.~\ref{figure_experiments_odd_viscosity}A.

\subsubsection{Compressible flows}

The effect of $\eta^o$ can be more dramatic in compressible flows. 
When odd viscosity is present in weakly compressible flows (at low Mach number), one should expect vorticity to induce density changes implied by $\delta \rho \propto \delta p = - \eta^o \omega$~\cite{Banerjee2017}.
For example, a Lamb–Oseen vortex without odd viscosity exhibits a density dip due to inertia. This dip can either be deepened or changed into a peak depending on the relative signs of odd viscosity and vorticity~\cite{Banerjee2017}.
In strongly compressible flows, odd viscosity can also lead to transverse flows under the influence of the term $\eta^{\rm o} \epsilonb \cdot \Delta \vec{v}$ in Eq.~\eqref{ns2d}, which is rotated by \ang{90}  with respect to $\Delta \vec{v}$.
This occurs for instance in compression shocks~\cite{Banerjee2017,Han2021}, as illustrated in figure \ref{figure_effects_odd_viscosity}H, in which a flow develops transverse to the direction of travel of the shock.

\subsubsection{Boundary effects}
\label{viscosity_boundary_effects}

Odd viscosity has an effect on the flow even in incompressible 2D fluids when boundary conditions involve the stress. For instance, surface waves have been used to measure odd viscosity in a colloidal chiral fluid made of spinning magnetic cubes~\cite{Soni2019}.  In these experiments, the antisymmetric part of the stress tensor, (see Box~\emph{Antisymmetric Stress}) corresponding to the terms $\tau$ and $\eta^{\rm R}$ in Eq.~(\ref{general_stress_2d}), produces chiral surface waves on top of a persistent particle current at the edge. While not responsible for the presence of these surface waves, odd viscosity modifies their dispersion relation and damping rate in a way that can be measured. 

In systems without antisymmetric stress, odd viscosity induces distinct chiral surface waves with dispersion $\Omega \approx - 2 \nu_{\text{o}} q |q|$ at a free surface (with a no-stress boundary condition) \cite{Abanov2018,Abanov2020,Bogatskiy2019}. 
Another class of boundary effects are topological sound waves~\cite{souslov2019topological,Tauber2019,Baardink2021}. These occur in compressible odd viscous flows, under the combined influence of the Lorentz (or Coriolis) body force and odd viscosity. We refer to \cite{Shankar2020,Fruchart2013} for a pedagogical introduction to topological waves; the gist is that these topological waves are unidirectional sound waves appearing at the boundary of the fluid because it has to untangle its internal topology (Fig.~\ref{figure_effects_odd_viscosity}A). Here, odd viscosity provides a short-distance cutoff that makes the topological invariant well-defined and triggers unusual topological phase transitions~\cite{souslov2019topological,Tauber2019}.

Finally, the drag and lift forces on an object embedded in a 2D fluid with odd (shear) viscosity have been analyzed in \cite{Kogan2016,Hosaka2021b,Lapa2014,Ganeshan2017,Lier2022}. 
The total lift force on a fixed object in a 2D incompressible fluid is not changed by $\eta^o$, although the contributions to the lift force from pressure and shear stress are both modified~\cite{Ganeshan2017}. A change in the lift force has however been reported in compressible odd fluids~\cite{Lier2022}.

\subsubsection{Low Reynolds number limit: Stokes flows}

In the limit of low Reynolds numbers\footnote{The dimensionless Reynolds number $\text{Re} = \rho U L/\eta$, where $U$ is a characteristic velocity and $L$ a characteristic length, measures the ratio between inertial and viscous forces. In some situations, it can be helpful to define by analogy an odd Reynolds number $\text{Re}^{\text{o}} = \rho U L/\eta^{\text{o}}$ (see for instance \cite{Banerjee2017,souslov2019topological}).}, the term $\vec{v} \cdot \nabla \vec{v}$ can be ignored in the Navier-Stokes equation \eqref{ns}, which reduces to the so-called Stokes equation. Since the Stokes equation is linear, the general solutions are expressed in terms of its Green function $G_{i j}(\vec{x})$, called the Oseen tensor, which gives the flow in response to a point force~\cite{KimKarrila,Happel2012}. 
In normal fluids, there is a symmetry in the exchange between the source (that produces a force) and the receiver (that measures the
velocity field) called Lorentz reciprocity, which is expressed by $G_{ij} = G_{ji}$~\cite[\S~4.2, Eq.~4.7]{Masoud2019}.
In fluids with odd viscosity, $G_{i j}(\vec{x}) \neq G_{j i}(\vec{x})$, so Lorentz reciprocity is broken~\cite{Hosaka2021,Khain2022}.

In \cite{Lapa2014}, swimming in a 2D fluid with odd (shear) viscosity is analyzed. The fluid produces a torque on the swimmer when it changes area. In three-dimensional Stokes flow (low Reynolds number), odd viscosity can no longer be absorbed in the pressure, even if the fluid is incompressible, leading to azimuthal flows that would otherwise be absent~\cite{Khain2022}. This is illustrated in the flow past a sphere in Fig.~\ref{figure_effects_odd_viscosity}D, where an azimuthal velocity appears with opposite sign above and below the sphere.

\subsubsection{Hydrodynamic instabilities}

Hydrodynamic instabilities are abrupt changes in a fluid flow that arise when a control parameter is varied~\cite{Charru2009}. How does odd viscosity affect these instabilities? 

Let us first ask a simple question: is an odd viscous fluid at rest linearly stable? It turns out that the answer is determined solely by the dissipative viscosities. To see that, let us linearize \eqref{ns} with \eqref{stress_strain_rate_relation} about the state $\vec{v} = \vec{0}$ with a uniform density $\rho$, yielding $\rho \partial_t v_i = - \eta_{i j k \ell} q_j q_\ell v_k$. Multiplying by $v_i$, we find
\begin{equation}
    \rho \partial_t \left( \frac{\lVert \vec{v} \rVert^2}{2} \right) = - \eta_{i j k \ell} q_j q_\ell v_k v_i = - \eta_{i j k \ell}^{\text{S}} q_j q_\ell v_k v_i
\end{equation}
in which the odd viscosities drop out in a similar way as in Eq.~\eqref{viscous_dissipation}. In particular, if the symmetric part of the matrices of viscosities is positive-definite, the fluid is linearly stable~\footnote{Formally, the case of odd elasticity with an overdamped dynamics is identical, see section \ref{sec_elastodynamics}.}.

\begin{figure}[h]
\includegraphics[width=14cm]{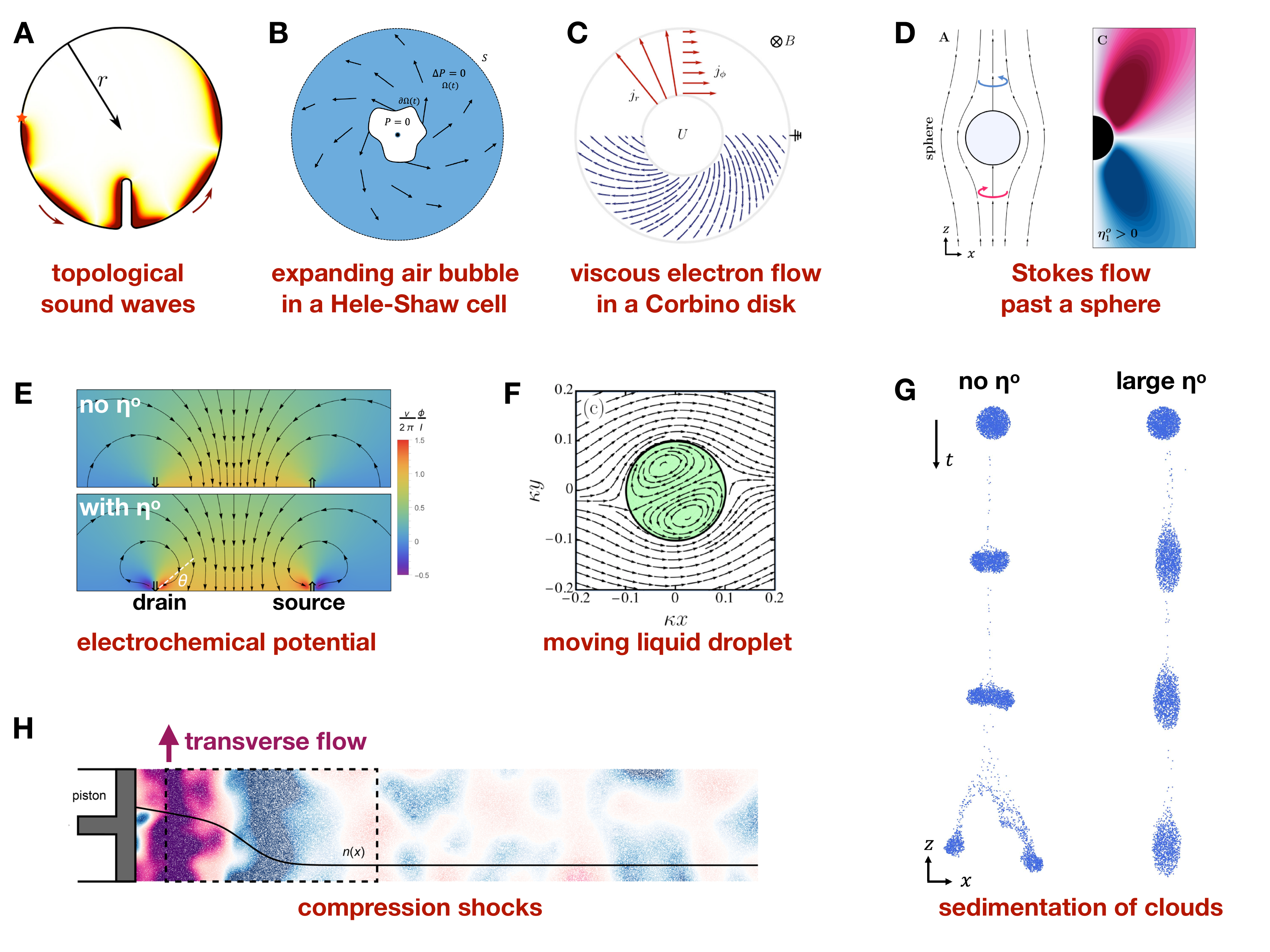}
\caption{\label{figure_effects_odd_viscosity}
\strong{Phenomenological consequences of odd viscosity.}
({\bf A}) Topological sound waves arise from a combination of the Coriolis/Lorentz force and odd viscosity. Adapted from \cite{souslov2019topological}.
({\bf B}) A chiral flow arises when a bubble expands inside a Hele-Shaw cell filled with a 3D odd viscous fluid. The flow acquires a circulation far from the bubble (represented by spiraling arrows) as a result of odd viscosities.  Adapted from \cite{Reynolds2021}.
({\bf C}) A chiral flow arises in a Corbino disk filled with an odd-viscous electron fluid under magnetic field.
The electric current (in blue) between the inner and the outer disks spirals, because it has both a longitudinal and a transverse component (in red) with a ratio $j_\phi/j_r = \eta^{\rm o}/\eta$.
Adapted from \cite{Holder2019}.
({\bf D}) The Stokes flow past a sphere develops an azimuthal component (blue and red arrows) in an odd viscous fluid Adapted from \cite{Khain2022}.
({\bf E}) Odd viscosity modifies the electrochemical potential near source and drain in inhomogeneous charge flows in solid-state systems. Adapted from \cite{Delacretaz2017}.
({\bf F}) A moving liquid droplet consisting of an odd viscous fluid (green) in a normal liquid exhibits a non-axisymmetric flow. Adapted from \cite{Hosaka2021b}.
({\bf G}) An instability in the sedimentation of clouds of particles is suppressed by odd viscosity. Adapted from \cite{Khain2022}.
({\bf H}) A transverse flow appears in a compression shock in an odd viscous fluid. Adapted from \cite{Han2021}.
}
\end{figure}

The effect of odd viscosity can be more dramatic in more complex flows, in which it can both lead to destabilization and stabilization. 
For instance, the magnetized Kelvin-Helmholtz instability (in magnetohydrodynamics) is modified by the presence of odd viscosities and is either stabilized or destabilized depending on the sign of the odd viscosity coefficient~\cite{Faganello2017,Nagano1979,Wolff1980}.
Odd viscosity also prevents the breakup of clouds of sedimenting particles \cite{Khain2022} in an odd viscous fluid (see figure \ref{figure_effects_odd_viscosity}G), and stabilizes the Saffman-Taylor instability~\cite{Reynolds2021}.
Further examples have been studied in falling thin films \cite{Lucas2014,Kirkinis2019,Bao2021,Chattopadhyay2021,Samanta2022} and in plasmas~\cite{Rosenbluth1962,Newcomb1966,Roberts1962} (more references in section \ref{sec_plasma}). 

\subsection{Statistical thermodynamics of odd viscosity}

The viscosity tensor represents the diffusive transport of momentum. 
Near equilibrium, the theory of transport processes is structured around two concepts~\cite{Pottier2010,Callen1985,DeGrootMazur}: 
the Green-Kubo relations provide a link between the transport coefficients and fluctuations at equilibrium; and the Onsager-Casimir reciprocity relations provides algebraic constraints on between transport coefficients. Neither of these principles are guaranteed in general far from equilibrium, raising the question to what extent they apply to odd viscosity.

\subsubsection{Green-Kubo Relations}
\label{paragraph_kubo}

In fluctuating hydrodynamics, the stress tensor fluctuates about its mean value. These fluctuations are characterized by the time correlation functions $\left\langle \sigma_{i j}(t) \sigma_{k \ell}(0) \right\rangle$ (we ignore spatial fluctuations for simplicity). When the correlations are symmetric in time, i.e. when 
\begin{equation}
    \label{correlation_detailed_balance}
\left\langle \sigma_{i j}(t) \sigma_{k \ell}(0) \right\rangle= \left\langle \sigma_{k \ell}(0) \sigma_{i j}(t) \right\rangle
\end{equation}
then we say that the system obeys \emph{detailed balance} or \emph{microscopic reversibility}. At equilibrium, the Green-Kubo formula states that the viscosities can be expressed in terms of these correlation functions as
\begin{equation}
  \label{kubo}
  \eta_{i j k \ell} = \frac{\mathcal{A}}{k_{\rm B} T} \int_{0}^{\infty} \left\langle \sigma_{i j}(t) \sigma_{k \ell}(0) \right\rangle dt
\end{equation}
in which $\mathcal{A}$ is the area (or volume in 3D) of the system. Fluids with odd viscosity can exhibit the same relation~\cite{Han2021,Epstein2020,Hargus2020,Fruchart2022}. In particular, inspection of Eqs.~\eqref{correlation_detailed_balance} and \eqref{kubo} reveals that odd viscosity arises when the correlation functions are not symmetric in time, namely when
\begin{equation}
    \left\langle \sigma_{i j}(t) \sigma_{k \ell}(0) \right\rangle \neq \left\langle \sigma_{k \ell}(0) \sigma_{i j}(t) \right\rangle.
\end{equation}

\begin{marginnote}[]
\entry{Green-Kubo relation}{a Green-Kubo formula provides a relation between the fluctuations of a system at rest and its response to a finite perturbation}
\end{marginnote}

\noindent Green-Kubo relations have been used to described the viscosity of quantum Hall fluids and other solid-state systems~\cite{Bradlyn2012,Read2011,Offertaler2019}.
In the context of active fluids, these relations were derived using the Mori-Zwanzig projection operator formalism and verified using molecular dynamics simulations in Ref~\cite{Han2021} for the shear part of the viscosity tensor of a 2D active fluid, in which time-antisymmetric parts of the correlation functions have indeed been observed.
They can also be obtained from the Onsager regression hypothesis~\cite{Epstein2020,Hargus2020}, as well as from kinetic theory~\cite{Fruchart2022}. 
We emphasize, however, that the validity of the Green-Kubo relation is not a priori not guaranteed in non-equilibrium systems.

\subsubsection{Onsager-Casimir reciprocity}
\label{sec_onsager_casimir}

The Onsager-Casimir reciprocity relations are constraints on transport coefficients, usually associated with microscopic reversibility~\cite{DeGrootMazur,Krommes1993,Coleman1960}. 
For the viscosity tensor, the Onsager-Casimir relations are usually expressed as
\begin{equation}
    \label{albert_einstein_was_wrong}
    \eta_{i j k \ell}(\vec{B}) = \eta_{k \ell i j}(-\vec{B})
\end{equation}
in which $\vec{B}$ symbolizes all external sources of time-reversal breaking (such as magnetic field or rotation). 
The Onsager relations are a particular case of Eq.~\eqref{albert_einstein_was_wrong} where the left- and right-hand sides describe the same system, so $\eta_{i j k \ell} = \eta_{k \ell i j}$.
Focusing on the shear viscosities in isotropic 2D systems (see Eq.~\eqref{general_stress_2d}),  Eq.~\eqref{albert_einstein_was_wrong} corresponds to $\eta(\vec{B}) = \eta(-\vec{B})$ and $\eta^{\text{o}}(\vec{B}) = - \eta^{\text{o}}(-\vec{B})$.
When the matrix in Eq.~(\ref{general_stress_2d}) is not symmetric Onsager reciprocity is said to be violated. This occurs whenever $\eta^{\text{o}} \neq 0$. For all systems known to the authors, Onsager-Casimir reciprocity appears to hold: 
the antisymmetric component of the viscosity tensor changes sign under time-reversal while the symmetric component remains unchanged.

\begin{marginnote}[]
\entry{Onsager reciprocity}{symmetry of the matrix of transport coefficients}
\end{marginnote}

\begin{marginnote}[]
\entry{Onsager-Casimir reciprocity}{symmetry between the matrix of transport coefficients at opposite values of the time-reversal breaking field $\vec{B}$ (rotation, magnetic field, etc.)}
\end{marginnote}

However, the range of applicability of the relation \eqref{albert_einstein_was_wrong} has not yet been fully delineated. 
Current numerical~\cite{Han2021,Hargus2020} and experimental~\cite{Berdyugin2019} results for the shear viscosities $\eta$ and $\eta^{\rm o}$ are in agreement with equation \eqref{albert_einstein_was_wrong}, which is also compatible with theoretical results from non-equilibrium thermodynamics~\cite{deGroot1954,DeGrootMazur} and from kinetic theory \cite{tenBosch1984,Sharipov1994,Sharipov1994b,Fruchart2022}\footnote{
When the Green-Kubo relations discussed in paragraph \ref{paragraph_kubo} apply, the Onsager relations arise as a consequence of detailed balance \eqref{correlation_detailed_balance}.
In the simplified kinetic theory discussed in paragraph \ref{odd_viscosity_from_collisions}, the Onsager-Casimir relations can be obtained from Eq.~\eqref{viscosity_main} by assuming that $L(B) = L^\dagger(-B)$ (this relation may then be derived from microscopic considerations in a given system).}.
However, theoretical arguments based on the Onsager regression hypothesis \cite{Hargus2020,Epstein2020} suggest that non-dissipative (odd) viscosities can exist in systems that do not break time-reversal symmetry at the level of stress correlations. 
In addition, the fact that the breaking of mirror symmetry and time-reversal invariance originate in the same physical phenomena in these systems can be a confounding factor, as the two symmetries impose constraints on the transport coefficients.

\begin{figure}[h]
\includegraphics[width=12cm]{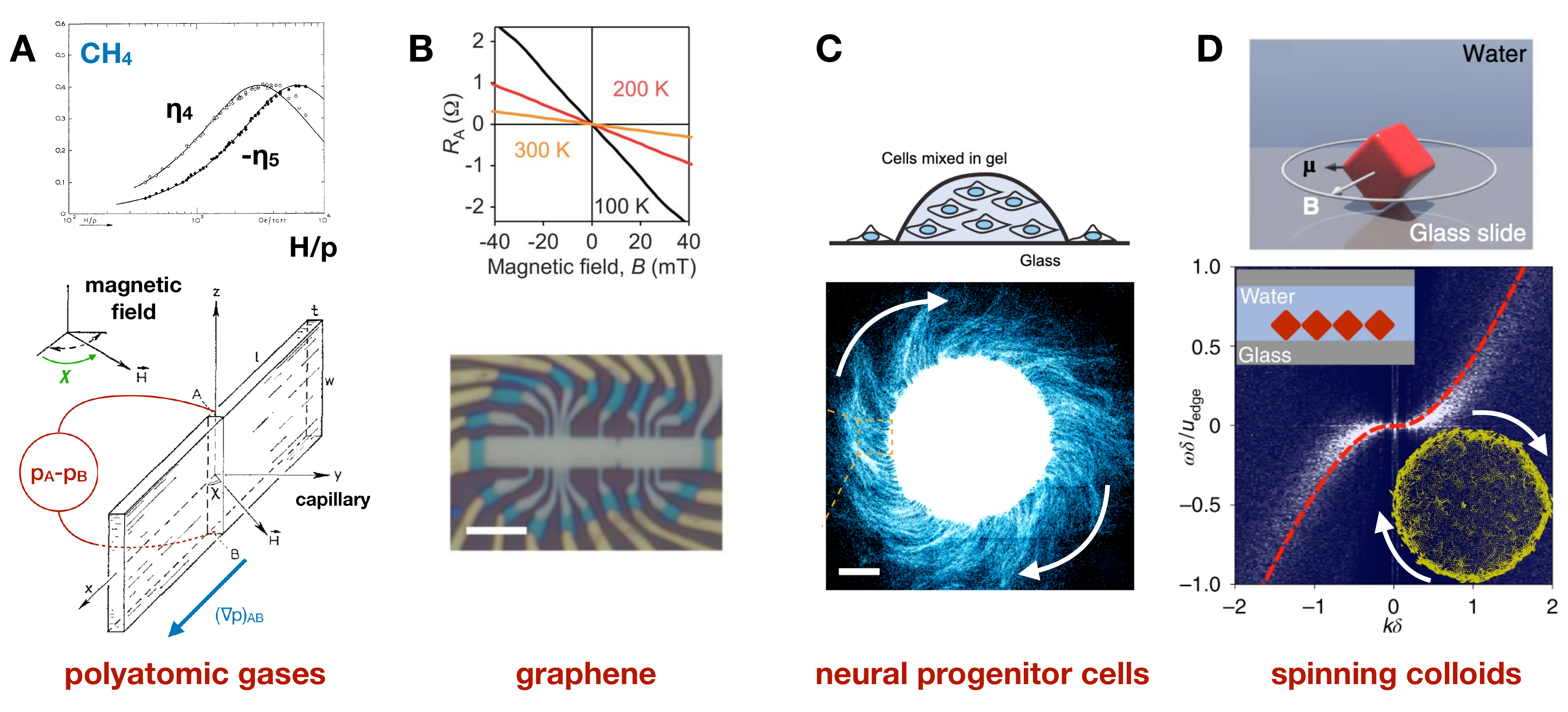}
\caption{\label{figure_experiments_odd_viscosity}
\strong{Experimental systems exhibiting odd viscosity.}
({\bf A}) Three-dimensional magnetized polyatomic gases such as $\text{N}_{2}$, $\text{CO}$, or $\text{CH}_4$ exhibit odd viscosities (called $\eta_4$ and $\eta_5$ in the nomenclature of \cite{DeGrootMazur}). When the gas flows in a capillary, a transverse pressure arises because of odd viscosities that allows for their measurement. As an example, the odd viscosities of $\text{CH}_4$ are plotted as a function of $H/p$, where $H$ is the external magnetic field and $p$ the overall pressure. Adapted from \cite{Hulsman1970b}.
({\bf B}) The two-dimensional fluid of electrons in graphene under a magnetic field exhibits a odd shear viscosity (Hall viscosity), which can be determined from multi-terminal electric transport measurements. Adapted from~\cite{Berdyugin2019}. 
({\bf C}) It has been proposed that the flow of in vitro neural progenitor cells can exhibit odd viscosity. Adapted from~\cite{Yamauchi2020}.
({\bf D}) The two-dimensional fluid consisting of small cubes with a permanent magnetic dipole moment in a colloidal suspension under a rotating magnetic field exhibits odd viscosity. It can be measured from the decay of surface waves originating from the antisymmetric part of the hydrostatic stress. Adapted from~\cite{Soni2019}. 
}
\end{figure}

\subsection{Where to find odd viscosity?}

According to the \enquote{central dogma} of phenomenology, anything that is not forbidden will, in principle, be present.
Odd viscosity can only occur when certain spatial (see sections \ref{2d_fluids} and \ref{3d_fluids}) and non-spatial (see section \ref{sec_onsager_casimir}) symmetries are broken.
For example, the shear odd viscosity $\eta^o$ can only exist if the fluid violates mirror symmetry.  
Below we discuss some systems that exhibit odd viscosity. In practice, all experimental systems we are aware of break time-reversal symmetry and mirror symmetry, and involve either spinning particles or an external magnetic field.

\subsubsection{Polyatomic gases under magnetic field}

Rarefied polyatomic gases (such as $N_2$ or $CO$) under a magnetic field exhibit transverse non-dissipative viscosities~\cite{Beenakker1970,Mccourt1990,Korving1966,Korving1967,Hulsman1970}, as well as transverse non-dissipative thermal conductivities (see Box~\emph{Other Oddities} and figure \ref{figure_experiments_odd_viscosity}A).
These effects have been observed in a variety of gases: we refer to \cite{Beenakker1970,Mccourt1990} for extensive reviews (see in particular Table 2 in \cite{Beenakker1970} and Table 7.1 in \cite{Mccourt1990} for lists of the various measurements known at the time). They originate from the precession of the magnetic moment of the particles due to the external magnetic field, which modifies the collision cross-section of the molecules.
These experimental studies are part of a large body of work in the '60-'70s on the effect of electric and magnetic fields on transport properties, that occurred hand in hand with progresses in kinetic theory~\cite{Beenakker1970,Mccourt1990}. The measurement of transverse viscosities has also been used to determine the sign of the so-called $g$-factor (a dimensionless proportionality constant relating the magnetic moment and the angular momentum of an object) \cite{Korving1966b}.

\subsubsection{Magnetized plasma}
\label{sec_plasma}

Electrically conducting fluids such as plasma or liquid metal can directly couple to a magnetic field (externally produced or induced by the fluid motion). These systems are described by a continuum theory called magnetohydrodynamics (MHD), in which the coupled dynamics of the fluid density $\rho$, its velocity field $\vec{v}$, and the magnetic field $\vec{B}$ are described. 
In a magnetized plasma, time-reversal symmetry and parity are broken, so we expect odd viscosities to occur. 
This is indeed what happens: the physical mechanism can be tracked down to the fact that charged particles in a magnetic field have a tendency to rotate along circular orbits because of the Lorentz force: this is reviewed in section \ref{sec_collisionless}.
In this context, odd viscosity is usually known as gyroviscosity. 

\begin{marginnote}[]
\entry{gyroviscosity}{synonym of odd viscosity in the context of plasma physics}
\end{marginnote}

The existence of gyroviscosity has been established from kinetic theory calculations~\cite{ChapmanCowling,Braginskii1958,Braginskii1965,Ramos2005} as well as from more heuristic approaches~\cite{Kaufman1960,Newcomb1966}. 
We refer to the discussions of the consequences in Ref.~\cite{Steinhauer2011} (in particular \S III.D.2; and references therein) in the context of magnetic confinement fusion and in Ref.~\cite{Stasiewicz1993,Faganello2017} (and references therein) in the context of the ionosphere and magnetosphere of planets. 
Gyroviscosity is expected to stabilize certain instabilities that would otherwise occur, and to destabilize otherwise stable situations, in qualitative agreement with experimental observations~\cite{Steinhauer2011,Nayyar1970,Steinhauer1990,Wolff1980,Terada2002,Rosenbluth1962,Newcomb1966,Roberts1962,Ishizawa2005,Ferraro2006,Bae2013,Stacey1985,Stacey2002,Stacey2006,Winske1996,Zhu2008,Yajima1966}. 
As an example, gyroviscous theories have been used to model the observed dusk-dawn asymmetry in the Kelvin-Helmholtz instabilities occurring in the magnetosphere of planets~\cite{Faganello2017,Wolff1980,Terada2002}.

\subsubsection{Gases under rotation}

Odd viscosity is expected to arise in gasses under rotation as a consequence of Coriolis forces. 
This prediction has been made using using kinetic theory in dilute gases~\cite{Nakagawa1956}, and we sketch a similar calculation in Sec.~\ref{sec_collisionless}. 
One finds that the the ratio of odd over normal viscosity is given by $\eta^{\text{o}}/\eta \sim \Omega \tau$, where $\Omega$ is the angular frequency of rotation and $\tau \sim \eta/p$ is a collision time, expressed in terms of the shear viscosity of the unperturbed fluid $\eta$ and the pressure $p$. 
This estimate suggests that obtaining a sizeable odd viscous response would require a combination of low pressures, high viscosities, and fast rotation. To give an order of magnitude, in order to have $\eta^{\rm o} \sim \eta$, the required spinning speed should be of the order of $\tau^{-1} \simeq \SI{1e9}{\per\second}$ for air at atmospheric pressure. 

\subsubsection{Condensed matter: graphene, superfluids, etc.}

The transport of electrons in a metal is usually described by Ohm's law, but this is not the only possibility~\cite{Sulpizio2019,Ku2020,Bandurin2016,Moll2016,Crossno2016,deJong1995,Zaanen2016,Lucas2018}. When only electron-electron collisions occur, the transport is hydrodynamic: electrons then behave like a viscous fluid\footnote{The diffusive behavior described by Ohm's law arises when only electron-lattice collisions are present. When all collisions are negligible, the transport is ballistic. When both electron-electron and electron-lattice collisions are present, the behavior of electrons is similar to the flow of a viscous fluid in a porous medium~\cite{Sulpizio2019}.}. This has been observed in high-mobility systems such as graphene~\cite{Sulpizio2019,Ku2020,Bandurin2016,Moll2016,Crossno2016,deJong1995}. In this regime, odd viscosities (called Hall viscosities in this context) occur when the electrons are put in a magnetic field, so time-reversal symmetry is broken. Hall viscosities lead to corrections to charge transport quantities such as resistances. An isotropic 2D shear Hall viscosity has been measured in graphene under magnetic field from transport measurement in multiterminal Hall bars~\cite{Berdyugin2019}, where the normal shear viscosity is $\nu \approx \SI{3e-4}{\meter^2\per\second}$ at $T \approx \SI{300}{\kelvin}$, while $\nu^{\text{o}} \approx \SI{20e-4}{\meter^2\per\second}$ at $B \sim \SI{40}{\milli\tesla}$. Here, $\nu=\eta/{\rho}$ and $\nu^{\text{o}}=\eta^{\text{o}}/\rho$.

\begin{marginnote}[]
\entry{Hall viscosity}{synonym of odd viscosity in the context of condensed matter physics}
\end{marginnote}

Odd viscosity is also expected in superfluids under magnetic field or under rotation such as liquid helium and other chiral superfluids~\cite{Vollhardt2013,Volovik2020,Furusawa2021,Fujii2018,Read2009,Read2011}. In gapped quantum Hall fluids, odd viscosity contains information about the topological order of the state~\cite{Avron1995,Haldane2009,Haldane2011,Park2014,Hoyos2012,Delacretaz2017,Avron1998,Read2009,Read2011,Bradlyn2012}. In this case, in a rotationally invariant 2D system, the Hall viscosity is proportional to a quantized topological invariant $S$ called shift through $\eta^{\text{o}} = \hbar ({S}/{4}) n$ where $\hbar$ is the Planck constant and $n$ the electron density.

\subsubsection{Active and driven soft matter} 
Fluids composed of active objects exhibiting a preferred chirality are called chiral active fluids~\cite{Furthauer2012,vanZuiden2016,Banerjee2017,Markovich2019,Han2021,Yeo2015}.
They include collections of self-spinning colloidal particles \cite{Grzybowski2000,Yan2015,Soni2019,Bililign2021}, driven chiral grains~\cite{Tsai2005}, 3D-printed rotors fluidized by a turbulent upflow~\cite{LopezCastano2021,LopezCastano2022},
as well as robotic systems~\cite{Scholz2018,Yang2020} and biological matter \cite{Riedel2005,Petroff2015,Yamauchi2020,Tan2021}.
Theoretical works have predicted that these systems exhibit odd viscosity as well as hydrostatic torques (see Box~\emph{Antisymmetric Stress})~\cite{Banerjee2017,Markovich2021}.

Experimental measurements of odd viscosity have been performed in a fluid made of spinning colloids~\cite{Soni2019}.
The colloidal cubes suspended in water over a glass surface posses a magnetic moment, and spin under the effect of a rotating magnetic field. Odd viscosity is then measured through modification in the dispersion of surface waves (see section \ref{viscosity_boundary_effects}), and the values reported are $\eta^{\rm o} = \SI[separate-uncertainty]{1.5(1)e-8}{\pascal\meter\second}$
and 
$\eta = \SI[separate-uncertainty]{4.9(2)e-8}{\pascal\meter\second}$.

Odd viscosity has also been numerically observed in simulations of dense but passive chiral fluids composed of ratchets in simple shear and planar extensional flows~\cite{Zhao2021,Zhao2022} (see also \cite{Garzo2017}). 
Magnetized nematic ferrofluids are also expected to exhibit odd viscosities~\cite{Hess1986,Jarkova2001}.

\subsubsection{Vortex matter}
In an ideal fluid, stable point vortices can exist and interact with each other. 
A large collection of such vortices can itself be treated as a fluid. 
It has been predicted that such a two-dimensional fluid composed of vortices of the same sign exhibits odd viscosity~\cite{wiegmann2014anomalous,Yu2017,Moroz2018}.
Intuitively, this can be understood from the fact that the vortices have transverse interactions (breaking the symmetries that would forbid odd viscosity in 2D), but form a Hamiltonian system (which suggests that normal viscosities must be absent).
It has also been reported that numerical simulations of skyrmions in chiral magnets exhibit odd viscosity \cite{Reichhardt2022}.

\subsection{Microscopic mechanisms}

Microscopically, odd viscosity can arise from at least two classes of mechanisms: (i) as a single-particle effect originating in individual particle's dynamics between collisions (this is discussed in section \ref{sec_collisionless}) and (ii) as an effect of interactions between particles (section \ref{odd_viscosity_from_collisions}).
We will first discuss the physical mechanisms, and then sketch how continuum equations can be obtained from microscopics using kinetic theory (section \ref{kinetic_theory}). Finally, we will sketch how odd viscosity can be obtained from variational principles consistent with microscopic symmetries (section \ref{action_principles}).

\begin{figure}[h]
\includegraphics[width=12cm]{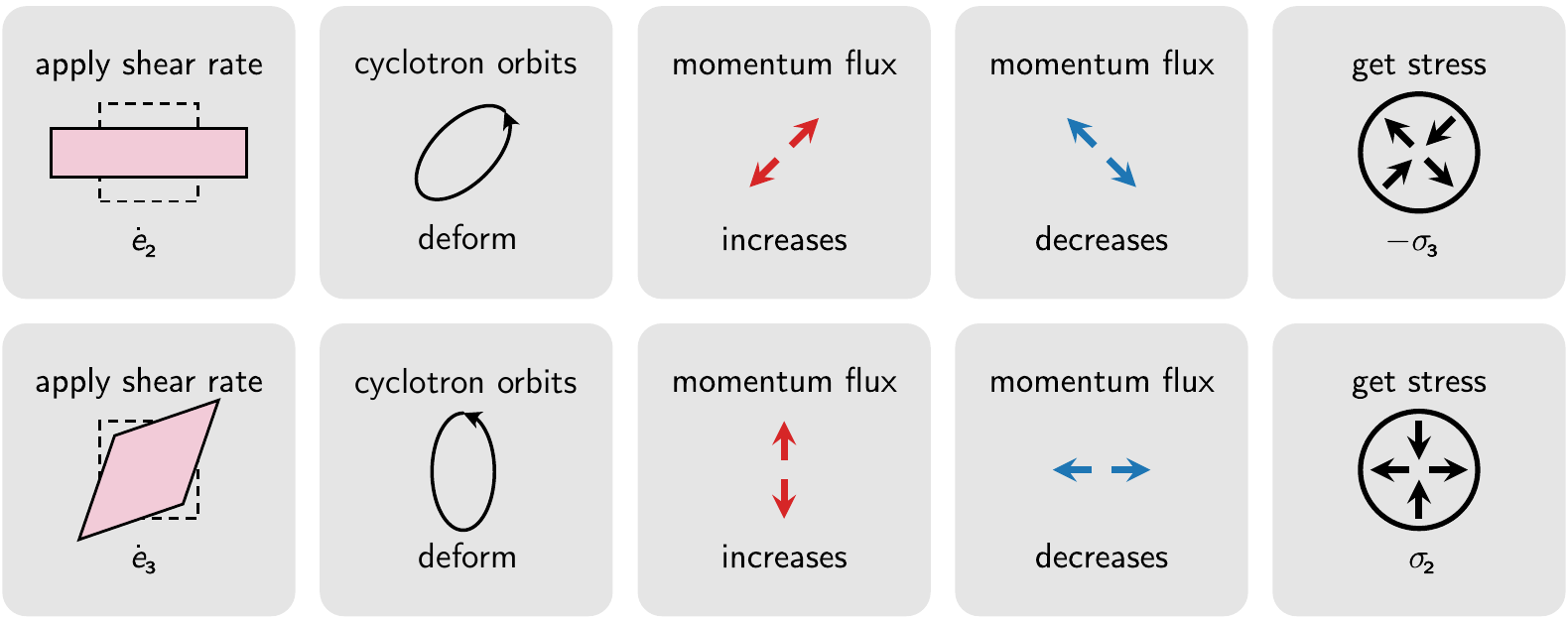}
\caption{\label{figure_viscosity_collisionless}
\textbf{Collisionless odd viscosity.}
In a magnetized plasma, the charged particles follow closed trajectories called cyclotron orbits. In a fluid at rest, the orbits are circular. When a shear rate (in the plane orthogonal to the magnetic field) is applied to the fluid, the cyclotron orbits become elliptic. The principal axes of the ellipse are rotated by \ang{45} with respect to the shear rate axes in a direction set by the magnetic field. This leads to a stress rotated by \ang{45} from the shear rate, corresponding to odd viscosity. 
}
\end{figure}

\subsubsection{Collisionless odd viscosity}
\label{sec_collisionless}

Odd viscosity can be generated by the individual motions of particles in an external field.
The Lorentz force (for charged particles in a magnetic field) and Coriolis forces (for all particles in a rotating reference frame) tend to move the particles along circular orbits. This single-particle effect leads to odd viscosity. 
To illustrate how this works, consider a gas of charged particles in an external magnetic field, as shown in Fig.~\ref{figure_viscosity_collisionless}.  
Recall that an individual charged particle in a magnetic field follows a circular orbit called Larmor orbit or cyclotron orbit, whose center is called the guiding center and radius the Larmor radius (see Refs.~\cite{Goldston2020,Hazeltine2018} for more about plasma physics). When the Larmor radius is small enough (in a thermal plasma, this occurs at large enough magnetic field), the gyrating particle behaves as a magnetic point dipole aligned with the field and located at the guiding center.

The guiding center of the particles in the plasma can move under the influence of external forces and of other particles. This motion is summarized by the velocity field of the plasma. 
When a velocity gradient is present, the cyclotron orbit deforms into an ellipse. This can be shown explicitly from the equations of motion for a single particle in a spatially varying force field~\cite{Kaufman1960,Newcomb1966}. As illustrated in Fig.~\ref{figure_viscosity_collisionless}, the ellipse is rotated by \ang{45} with respect to the velocity gradient. The resulting momentum flux is therefore also rotated by \ang{45}. This means that a shear rate 1 in Table~\ref{table_ir_2d} leads to a shear stress 2, while a shear rate 2 leads to minus a shear stress 1: this corresponds to the odd shear viscosity $\eta^{\rm o}$ in Eq.~\eqref{general_stress_2d}.
This effect is due to the deformation of the Larmor orbits. Because it goes beyond the picture of effective dipoles, it is known as a finite Larmor radius effect, or as a non-ideal MHD effect.
This effect can be captured in a more systematic manner using kinetic theory: we refer to \cite{ChapmanCowling,Braginskii1958,Braginskii1965,Kaufman1960,Newcomb1966} for more details.

\subsubsection{Odd viscosity from collisions}
\label{odd_viscosity_from_collisions}

The preceding section discussed a mechanism in which odd viscosity is essentially a single body effect: the free trajectory of a particle is modified in between collisions. However, odd viscosity can also arise as the result of abnormal collisions, even when the free path is unchanged. 
In typical fluids, the collisions are, on average, invariant under mirror reflections (Fig.~\ref{figure_viscosity_collisions}AB). 
However, collisions that on average violate parity (i.e. mirror reflection symmetry, as in Fig.~\ref{figure_viscosity_collisions}CD)~\cite{Soni2019,Bililign2021,Tsai2005} can lead to odd viscosity~\cite{Fruchart2022,Han2021}. The mechanism can be understood pictorially (Fig.~\ref{figure_viscosity_collisions}).
When we subject the fluid to a constant shear rate shown in panel E, vertical collisions are more frequent and horizontal collisions are less frequent than in the fluid at rest.
The resulting change in the momentum flux is qualitatively obtained by looking at where particles go after collision.
When the collisions are asymmetric (panel D), there is an increase of the momentum flux at \ang{45} and a decrease of the momentum flux at \ang{-45}.
The corresponding viscous stress is the opposite of this momentum flux.
Repeating this argument for a constant shear rate at \ang{45} (panel F), we find an increase in the vertical momentum flux and a decrease in the horizontal momentum flux. The relations between the resulting stresses and the strain rates we apply in E and F are antisymmetric: this is odd viscosity. 
Realistic microscopic descriptions of magnetized neutral polyatomic gases, which take into account internal molecular degrees of freedom, indeed agree quantitatively with measured non-dissipative transport coefficients~\cite{Korving1966,Korving1967,Kagan1962,Kagan1962b,Kagan1967,Moraal1969,McCourt1969,Hulsman1970,Knaap1967,McCourt1967,Levi1968,Waldmann1958b,Hess2003,Beenakker1970,Mccourt1990}. The main ideas from these calculations can be captured via a simplified model in which the internal degrees of freedom are neglected and only the fact that collisions are not invariant under mirror symmetry are kept~\cite{Fruchart2022}. 

\subsubsection{Kinetic theory}
\label{kinetic_theory}

\begin{figure}[h]
\includegraphics[width=15cm]{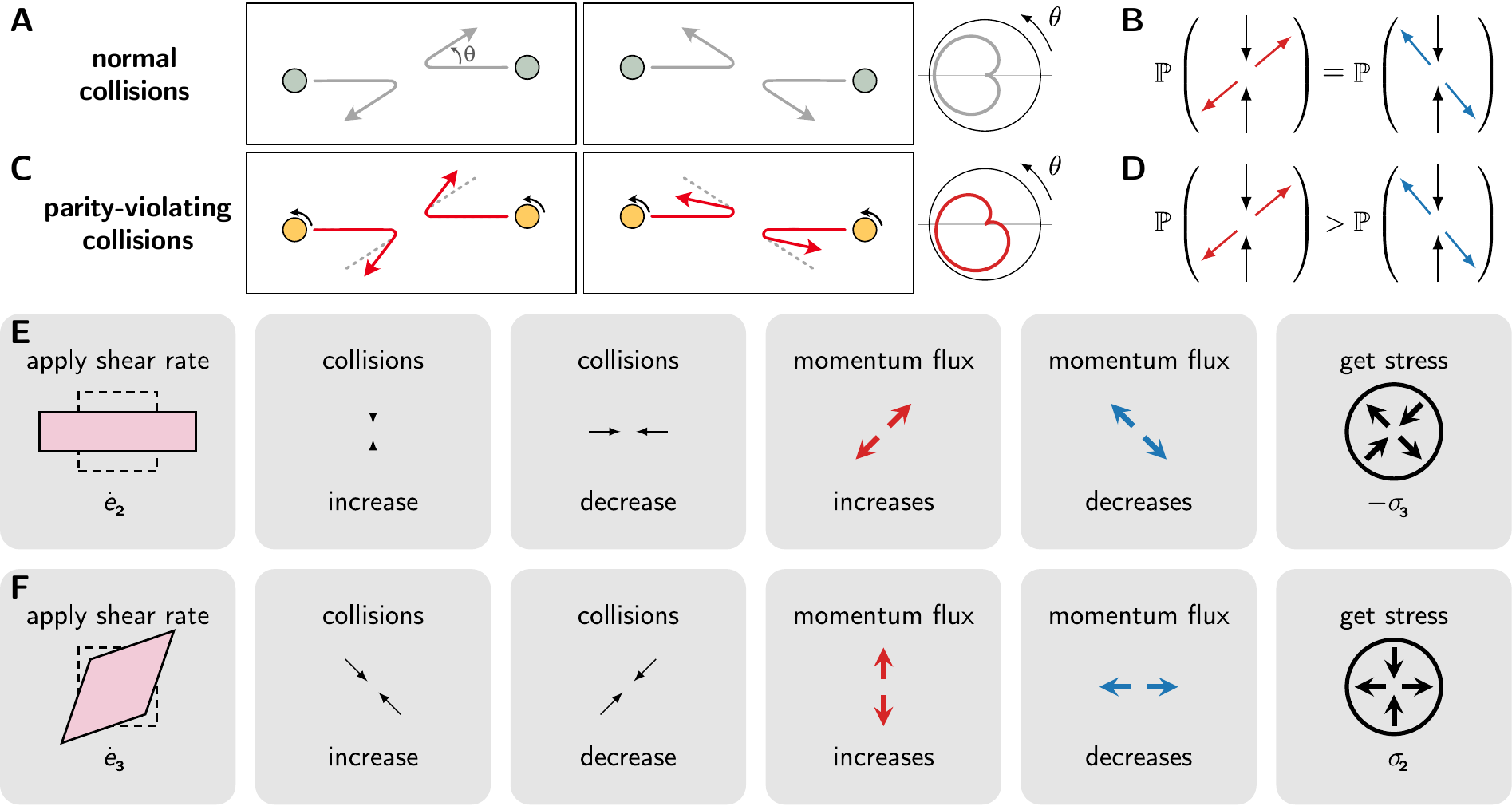}
\caption{\label{figure_viscosity_collisions}
\textbf{Odd viscosity from collisions.}
In the case of normal collisions ({\bf A}), the cross-section $\sigma(g, \theta)$ giving the probability ({\bf B}) of particles going out with an angle $\theta$ is symmetric: $\sigma(g, \theta) = \sigma(g, -\theta)$, where $g$ is the center of mass momentum. ({\bf C-D}) This is not the case for parity-violating collisions, for which $\sigma(g, \theta) \neq \sigma(g, -\theta)$. 
In line ({\bf E}), the fluid is subjected to a constant shear rate $\dot{e}_{2}$ (see Table~\ref{table_ir_2d}). 
As a consequence, there are more vertical collisions and less horizontal collisions, compared to the fluid at rest.
The change in the momentum flux (compared to the fluid at rest) is qualitatively obtained by looking at where particles go after collision.
As the collisions are asymmetric (see panel D), there is an increase of the momentum flux at \ang{45} (in red) and a decrease of the momentum flux at \ang{-45} (in blue).
Combining these, we find the resulting viscous stress $-\sigma_{3}$ in the last panel.
Note that the momentum flux tensor (pressure tensor) is the opposite of the stress tensor.
There is also a change in the horizontal and vertical momentum fluxes, not pictured there. It corresponds to normal shear viscosity. 
In ({\bf F}), we follow the same reasoning as in D when the the fluid is subjected to a constant shear rate $\dot{e}_{3}$. 
The result is a viscous stress $\sigma_{2}$.
Hence, we have found that $\sigma_{2} \propto \dot{e}_{3}$ and $\sigma_{3} \propto - \dot{e}_{2}$ (ignoring normal shear viscosity), which is indeed the effect of odd viscosity.
Adapted from \cite{Fruchart2022}.
}
\end{figure}

In this section, we sketch how (odd) viscosities can be obtained from microscopic models using kinetic theory. We focus on the simplest cases, and refer the reader to the literature~\cite{Korving1966,Korving1967,Kagan1962,Kagan1962b,Kagan1967,Moraal1969,McCourt1969,Hulsman1970,Knaap1967,McCourt1967,Levi1968,Waldmann1958b,Hess2003,Beenakker1970,Mccourt1990} for more realistic cases.
To do so, let us consider the distribution function $f(t,\vec{r},\vec{c})$, giving the probability $f(t,\vec{r},\vec{c}) \dd^d \vec{r} \dd^d \vec{c}$ to find a particle in a volume centered at position $\vec{r}$ and velocity $\vec{c}$ in phase space at time $t$. The fluid at rest is described by a stationary distribution such as the Boltzmann distribution $f^\circ \propto \exp[ \lVert\vec{c} - \vec{v}\rVert^2/(2 k_{\text{B}} T)]$ ($\vec{v}$ is the fluid velocity, $T$ the temperature, $m$ the particles' mass, $k_{\text{B}}$ the Boltzmann constant).
Viscosity is contained in the way the distribution function relaxes towards equilibrium after being perturbed by a velocity gradient.
This relaxation can be described by kinetic theories, such as the Boltzmann equation which linearized about the stationary distribution reads
\begin{equation}
	\label{LBE_main}
	\frac{\partial \phi}{\partial t} 
	+ c_i \frac{\partial \phi}{\partial r_i}
	+ b_i \frac{\partial \phi}{\partial c_i}
	= L \phi
\end{equation}
where $f = f^\circ (1 + \phi)$, $m \vec{b}$ are body forces and $L$ is the linearized collision operator which can be expressed in terms of the scattering cross-section $\sigma(g, \theta)$ defined in Fig.~\ref{figure_viscosity_collisions}. Equation~(\ref{LBE_main}) expresses the conservation probability, where the left-hand side captures the probability flow without interactions and $L$ on the right-hand side captures the redistribution of probability due to collisions.  

Let us first consider the case of chiral collisions with no external force ($\vec{b} = \vec{0}$). 
The viscosity tensor can be expressed as an inner product of the form~\cite{Fruchart2022}
\begin{align}
	\label{viscosity_main}
 	\frac{\eta_{i j k \ell}}{\rho} &= - \frac{m}{k_{\text{B}} T} \, 
 	\left(c_i c_j, L^{-1} \, c_k c_\ell\right)
\end{align}
An explicit calculation shows that when collisions are chiral ($\sigma(g, \theta) \neq \sigma(g, -\theta)$), the linearized collision operator $L$ becomes non-Hermitian ($L \neq L^\dagger$), which leads to odd viscosity $\eta_{ijk\ell}\neq\eta_{k\ell ij}$ through Eq.~(\ref{viscosity_main}).

The case of particles in a magnetic field can be treated in a similar way. The charged particles are subject to an out-of-plane magnetic field $B$ and experience a Lorentz force $b_i(\vec{c}) = \omega_{\text{B}} \epsilon_{i j} c_j$ where $\omega_{\text{B}} = (q/m) B$ ($q$ is the charge), or equivalently a Coriolis force on neutral particles in a frame of reference rotating at angular frequency $\Omega = \omega_{\text{B}}/2$.
Here, the collisions are not chiral, so we can model them using the so-called relaxation time approximation where $L \phi \simeq -\phi/\tau$ in Eq.~\eqref{LBE_main}, in which $\tau$ is a relaxation time.
The viscosity can then be obtained. A simplified calculation is given in Ref.~\cite{Kaufman1960}, in which we find that
$\eta = {p \tau}/[{1+ 4 \tau^2 \omega_{\text{B}}^2}]$ and 
$\eta^{\rm o} = 2 \omega_{\text{B}} \tau \, \eta$.
We refer to \cite{ChapmanCowling,Braginskii1958,Braginskii1965} for more detailed calculations.

\subsubsection{Action principles and Hamiltonian structure}
\label{action_principles}

The dynamics of an ideal fluid can be obtained from an action or a Hamiltonian~\cite{Morrison1998}.
The action of an ideal fluid is
\begin{equation}
    \label{action_ideal_fluid}
    S = \int \left[ \frac{1}{2} \rho \lVert \vec{v}\rVert^2 - \rho U \right] \, dt \, d^dr
\end{equation}
in which $\vec{v}$ is the velocity field, $\rho$ the density field, and $U$ the internal energy per unit mass of the fluid. Imposing a stationary action $\delta S = 0$ directly leads to the Euler equation.
To be precise, the action in Eq.~(\ref{action_ideal_fluid}) is written in Eularian coordinates, but it must be varied in Lagrangian coordinates~\cite{Morrison1998}. From the action principle, one can deduce a Hamiltonian and a Poisson bracket structure as well. 
This point of view provides useful insights on conservation laws, the stability of the fluid, and approximation methods~\cite{Morrison1998,Salmon1988}.

Since odd viscosity does not contribute to dissipation, it is a form of ideal flow, and can be captured in this framework. 
The main idea, originating in the physics of plasma~\cite{Morrison1984,Lingam2014,Morrison2005,Morrison2014,Lingam2020,Lingam2015} and recently analyzed in the context of active and electronic matter \cite{Banerjee2017,Markovich2021,Monteiro2021}, consists in adding to the action a term
\begin{equation}
    \Delta S = \int \vec{L} \cdot \vec{\omega} \, dt \, d^dr
\end{equation}
in which $\vec{\omega} = \nabla \times \vec{v}$ is the vorticity, and $\vec{L}$ is a new quantity introduced phenomenologically, that sometimes corresponds to the density of angular momentum. 
In the context of gapped quantum Hall fluids, odd viscosity has been described by so-called topological terms known as the Wen-Zee and (gravitational) Chern-Simons actions~\cite{Wen1992,Avron1995,Hoyos2012,Abanov2014,Hughes2013,Gromov2016}. In the case of gapless fluids, it has been described by another kind of topological term known as the Wess-Zumino-Witten action~\cite{Geracie2014} (see also \cite{Haehl2015,Jensen2012} for other effective field theory approaches).

\begin{textbox}[h]\section{Other Oddities}
Odd responses similar to odd viscosity or elasticity arise in various contexts, often associated with a transverse response~\cite{Barabanov2015}.
Perhaps the most well-known is the Hall effect: a transverse voltage difference arises in response to a longitudinal electric current in electric conductors under magnetic field. This originates from the local Ohm law $\vec{j} = \vec{\sigma} \vec{E}$ relating the current density to the electric field, in which the electrical conductivity $\sigma_{i j}$ acquires an antisymmetric part called Hall conductivity.
Similar phenomena arises in the heat diffusion (under the name of thermal Hall or Righi-Leduc effect~\cite{Strohm2005}), in particle diffusion (see Refs.~\cite{Koch1987,Auriault2010,Hargus2021,LopezCastano2022} and references therein), in light diffusion~\cite{Rikken1996,vanTiggelen1995}, or in the case of Willis couplings~\cite{Quan2021}. 
We also refer to \cite{Barabanov2015} for a review of Hall-like effects.
Analogues of odd elasticity also occur in 
the generalized elasticity~\cite{Chaikin1995,Anderson1984}  of mesophases with nonvariational dynamics such as active cholesterics and hexatics~\cite{Kole2021,Maitra2020Chiral}.
As an example, consider the diffusion equation
\begin{equation}
    \partial_t \rho + \nabla \cdot \, \vec{j} = 0
    \quad
    \text{where}
    \quad
    \vec{j} = - {\vec D} \, \nabla \rho
\end{equation}
in which $\rho$ is the particle density. The antisymmetric part of the diffusion tensor $D_{i j}$ contains odd (Hall) diffusivities.
In a uniform system, $\nabla \cdot \, \vec{j} = D_{i j} \partial_i \partial_j \rho$, so odd diffusivities drop out of the equation of motion. They can still be observed provided that they enter the boundary conditions, or if the current can measured directly.
\end{textbox}

\section{Odd elasticity}

The canonical formulation of elasticity typically assumes that the stress strain relationship is compatible with a potential energy. This assumption is in general not appropriate for a range of living, driven, or active media, and odd elasticity is the enhanced theory that emerges when the assumption is removed.

\subsection{What is elasticity?}

When you gently deform a solid, such as a rubber pencil eraser, the solid resists the shape change you are trying to impart. This is due to elasticity, which captures the resistance of a solid to having inhomogeneities in its displacement field. The elastic solid exerts stresses captured by the stress tensor
\begin{equation}
    \sigma_{i j} = C_{i j k \ell} \partial_\ell u_k
\end{equation}
in response to shape changes, described by gradients of the displacement field $\vec{u}(t,\vec{x})$. The stress tensor describes the forces that a parcel of elastic continuum applies to the neighboring parcels, or to the environment. The coefficient of proportionality $C_{i j k \ell}$ is called the elasticity tensor, or elastic modulus tensor.

It is convenient to describe the deformation of an elastic medium with respect to a fixed undeformed reference state.
When a material is deformed, a point originally at position $\vec{x}$ is moved to a new location $\vec{X} (\vec{x})$, allowing us to define a displacement field $\vec{u}(\vec{x}) = \vec{X}(\vec{x}) - \vec{x}$. 
Fields expressed in terms of $\vec{x}$ are said to be in Lagrangian coordinates, which are different from the so-called Eulerian coordinates used in Section 2, in which fields are expressed as functions of points in the lab, without reference to an undeformed state. 
Depending on whether Lagrangian or Eulerian coordinates are used, the stress tensor comes in different flavors, respectively called the Piola-Kirchhoff (Lagrangian) stress $P_{i j}$ and Cauchy (Eulerian) stress $\sigma_{i j}$. 
Readers unfamiliar with the Piola-Kirchhoff stress tensor can ignore the distinction between 
$\sigma_{ij}$ and $P_{ij}$ in the following formulas as long as the pre-stress $P^{\text{(pre)}}_{ij}$ is zero\footnote{The Piola-Kirchhoff and Cauchy stresses are different, but they contain the same physical data~\cite{zubov2008nonlinear,Marsden1994}. They are related by $ P_{ik}J_{jk}= \det J  \sigma_{ij}$, where $J_{ij}= \pdv{X_i}{x_j} = \delta_{ij}+\partial_j u_i$ is the Jacobian of the map $\vec{X}$.
When there is no pre-stress, $P_{i j} = \sigma_{i j} + \mathcal{O}(\partial_i u_j)^2$ at first order in displacement gradients.
Please note that in this section, $\partial_i = \partial/\partial x_i$ where $x_i$ is a position in the undeformed reference state. In the section 2, $\partial_i = \partial/\partial X_i$ where $X_i$ is a fixed (Eulerian) position in the lab. See the S.I. for further discussion.
}.  
In linear elasticity, the stress is expressed as
\begin{align}
    P_{ij} = P^\pre_{ij}+ C^{\rm PK}_{ijk\ell} \, \partial_\ell u_k \label{eq:elasticity}
\end{align}
where $C^{\rm PK}_{ijk\ell}$ is a linear response matrix known as the elastic modulus tensor and $P^\pre_{ij}$ is the pre-stress present in the undeformed state, and $\partial_i u_j$ are displacement gradients.
We have assumed that the stress is a function of displacement gradients $\partial_i u_j$, so that solid-body translations do not cause any stress.

The elastic forces are given by $f_i = \partial_j P_{ij}$, and the power exerted by them is given by
\begin{align}
    \dot{W} &= \int f_i \dot{u}_i \dd^d x 
    = \int (\partial_j P_{ij}) \dot{u}_i \dd^d x 
    = \int (\partial_j P^\pre_{ij} + {C}^{\rm PK}_{i j k \ell} \partial_j \partial_\ell u_k) \dot{u}_i \dd^d x 
\end{align}
Upon performing an integration by part, we find that the elastic part of the power is 
\begin{align}
    \label{dotWel_intermediate}
    \dot{W}^{\text{el}} &= - \int {C}^{\rm PK}_{i j k \ell} (\partial_j \dot{u}_i) (\partial_\ell u_k)  \dd^d x 
\end{align}
in which we have ignored boundary terms and assumed that the elastic tensor ${C}^{\rm PK}_{i j k \ell}$ is uniform. We now compute the total work done during a cyclic evolution in time, during which the material is deformed, but goes back to its initial state at the end. It is
\begin{align}
    \Delta W^{\text{el}} &= \int \dot{W}^{\text{el}} \dd t
    = \int \dd t \, \dd^d x \, {C}^{\rm PK}_{i j k \ell} (\partial_j \dot{u}_i) (\partial_\ell u_k)
    = - \int \dd t \, \dd^d x \, {C}^{\rm PK}_{i j k \ell} (\partial_j u_i) (\partial_\ell \dot{u}_k)
\end{align}
in which the last equality is obtained from an integration by parts in time (note that the use of Lagrangian coordinates, and therefore of the Piola-Kirchhoff stress tensor, is necessary to perform this step).
After relabelling indices and summing the two equalities in \eqref{dotWel_intermediate}, we end up with
\begin{align}
    \label{Delta_W_el}
    \Delta W^{\text{el}} &= \int \dot{W}^{\text{el}} \dd t
    = - \int \dd t \, \dd^d x \, {C}^{\text{PK,A}}_{i j k \ell} (\partial_j \dot{u}_i) (\partial_\ell u_k)
\end{align}
in which only the antisymmetric part
\begin{equation}
    {C}^{\rm PK, A}_{i j k \ell} = [{C}^{\rm PK}_{i j k \ell} - {C}^{\rm PK}_{k \ell i j}]/2
    \qquad
    \text{(odd elastic tensor)}
\end{equation}
of the elasticity tensor appears.
As a consequence, the work performed by elastic forces over a cyclic deformation vanishes provided that the elastic tensor is symmetric (${C}^{\rm PK,A}_{i j k \ell} = 0$). 
The symmetry of the elasticity tensor ${C}^{\rm PK}_{i j k \ell}$ is known as Maxwell-Betti reciprocity~\cite{Truesdell1963Meaning}. 
An elastic medium where ${C}^{\rm PK,A}_{i j k \ell} \neq 0$ is called odd elastic, and violates Maxwell-Betti reciprocity. In this case, there is always some cyclic deformation such that $\Delta W^{\text{el}} \neq 0$ \cite{Scheibner2020}. 
This result is known as Betti's theorem~\cite{Truesdell1963Meaning}. 
More generally, the total work done by an odd elastic medium during a displacement depends on the path taken (not only on the end points of the trajectory in displacement space).
The linear response coefficient $C^{\rm PK}_{ijk\ell}$ is symmetric if an only if there is a potential $V$ such that $P_{i j} = {\partial V}/{\partial(\partial_j u_i)}$.

\begin{marginnote}[]
\entry{Odd elastic modulus}{an antisymmetric component of the elastic modulus tensor}
\entry{Maxwell-Betti reciprocity}{symmetry of the linear response matrix ${C}^{\rm PK}_{i j k \ell}$ relating forces and displacements}
\end{marginnote}

In Eq.~(\ref{eq:elasticity}), we needed the Piola-Kirchhoff stress in order to obtain valid expressions for the work $\Delta W$. 
In many situations, though, the stress tensor of interest is the Cauchy stress $\sigma_{ij}$. The Cauchy stress is advantageous because it does not require reference to the undeformed material, and it is symmetric when angular momentum is conserved (see Box~\emph{Antisymmetric Stress}). Just as with the Piola-Kirchhoff stress, we may expand the Cauchy stress as
\begin{align}
    \sigma_{ij} = \sigma_{ij}^\pre + C_{ijk\ell}\partial_\ell u_k
\end{align}
where $\sigma_{ij}^\pre$ is the (Cauchy) pre-stress present even in the undeformed medium. Notice that $C_{ijk\ell}$ is different from $C^{\text{PK}}_{ijk\ell}$.
As shown in the S.I., they are related to linear order in strain by
\begin{align}
    \label{PK_C_elastic_tensors}
    C_{ijk\ell}^{\text{PK}} = C_{ijk\ell} + \sigma^\pre_{ij} \delta_{k\ell} -\sigma^\pre_{i\ell} \delta_{jk}
\end{align}
Each independent component of $C_{ijk\ell}$ is referred to as an elastic modulus.

In our description, we have implicitly assumed that the stress tensor $\sigma_{i j}$ (or $P_{i j}$) is the current of linear momentum in the elastic continuum, as it is the case in usual elastic media. In this case, Eq.~\eqref{Delta_W_el} shows that a system with odd elasticity must have a mechanism for releasing and absorbing energy.  
The same formalism also describes systems in which $\sigma_{i j}$ does not represent a flux of momentum: $\vec{f}$ does not represent a force\footnote{It can be seen as a generalized force, but not a Newtonian force that could push a car.}, and $W$ does not represent an energy. These systems include coupled gyroscopes, vortices, and particles interacting through hydrodynamic interaction, that are discussed in section \ref{skyrmions_vortices_gyroscopes}.

\begin{figure*}[h]
\includegraphics[width=15cm]{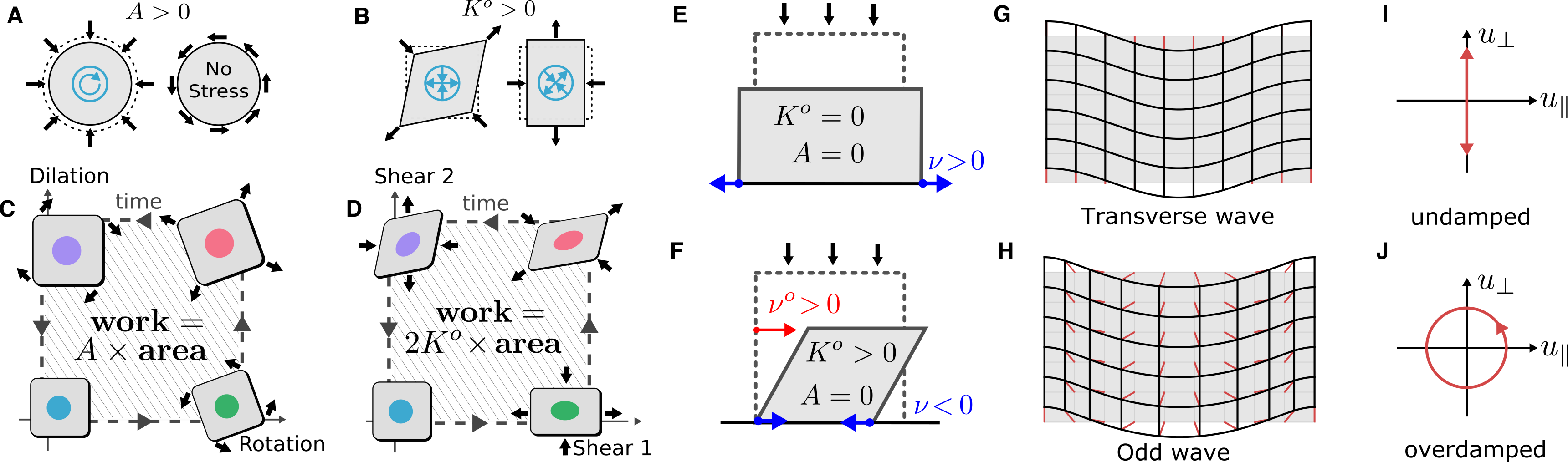}
\caption{\label{fig:elastic_casestudy}
{\bf Odd elasticity.} 
({\bf A}) The odd modulus $A$ induces a torque density when dilated or compressed. 
({\bf B}) The odd modulus $K^o$ induces a shear stress that is rotated with respect to the applied shear strain. 
The black arrows are a proxy for deformation, and the blue arrows indicate the stress. 
({\bf C-D}) Both $A$ and $K^o$ are nonconservative. This means that there exists (quasistatic) cycles of deformation along which the work done is nonzero.  
({\bf E}) A uniaxial compression of a passive material will generically result in symmetric deformation. If the material sides move outward, its Poisson ratio $\nu$ is positive. 
({\bf F}) A uniaxial compression of an odd elastic block results in a tilt whose intensity is proportional to the odd ratio $\nu^o$. If $K^o$ is sufficiently large, the material will also become auxetic ($\nu<0$), meaning that the material contract in response to compression. 
({\bf G}) A transverse wave in a passive solid is illustrated. 
({\bf H}) An odd elastic wave propagates with a circular polarization. The thin red lines represent the displacement field. 
({\bf I}) For an undamped inertial solid, the displacement field of a single plane wave oscillates back and forth along straight lines. 
({\bf J}) For an overdamped, odd elastic solid with $B=\mu=0$, the displacement field traces out closed ellipses. Adapted from~\cite{Scheibner2020}.
}
\end{figure*}

\begin{textbox}[h]\section{Nonvariational Dynamics of Order Parameters}
In systems with spontaneously broken symmetries such as critical phenomena and pattern formation, the dynamics of the order parameter $X$ can be modeled by a dynamical system $\partial_t X = f(X)$.
Such a dynamical system is called nonvariational (or non-potential) when it is not possible to express $f$ (locally) as the gradient of a potential ($f_i \neq - \partial_i V$) \cite{Fruchart2021,Nardini2017,Wittkowski2014,Kozyreff2007,Coullet1990,Coullet1989,Pomeau1983}. 
In this case, the Jacobian $J_{i j} = \partial_i f_j \neq J_{j i}$ is not symmetric.
In contrast, for potential systems, it must be symmetric 
$J_{i j} = - \partial_i \partial_j V = J_{j i}$.
This is reminiscent of odd elasticity and odd viscosity, except that the antisymmetry enters at the level of the linearized equation of motion rather than the constitutive relations. 
The non-variational nature of the dynamics leads to various effects that would not occur in purely relaxational systems, including 
time-dependent states with complex spatiotemporal structure, some of which have been experimentally observed~\cite{Bodenschatz1992,Clerc2005}.
These include rotating spiral states~\cite{Bodenschatz1992},
self-propelled localized structures such as dislocations, Bloch walls, and defects moving at constant speed~\cite{Pomeau1983,Siggia1981,Coullet1990,Colinet2002,Tsimring1996,Kozyreff2007,Clerc2005,Houghton2011},
proliferation of defects \cite{Colinet2002,Tsimring1996},
localized pulses \cite{Bodenschatz1992},
spontaneous parity breaking and traveling states~\cite{Coullet1989,Fruchart2021,You2020,Saha2020,FrohoffHulsmann2022},
and spatiotemporal chaos \cite{Clerc2013,Coullet1992,Bodenschatz1992,Fruchart2021}.
When a noise term is added to the dynamical system, the nonvariational nature of the dynamics is typically associated with  
broken time-reversal invariance, a nonvanishing rate of entropy production in the steady-state~\cite{Nardini2017,Wittkowski2014,Cates2019}, and non-reciprocal couplings or interactions \cite{Loos2019,Loos2020}.
\end{textbox}

\subsection{Case study: 2D isotropic media}

To illustrate odd elasticity in a concrete setting, we ask: what does odd elasticity look like in a two dimensional isotropic solid? Besides providing a mathematically simple illustration of the general concepts~\cite{Scheibner2020}, this setting is relevant for many of the experimental systems discussed in section~\ref{sec:micromodels}.

\subsubsection{Two-dimensional odd elastic moduli} 
Since $C_{ijk\ell}$ is a rank 4 tensor, in two dimensions it has at most $2^4=16$ independent coefficients. However, as we will see, assumptions such as spatial symmetries, coupling to rotations, angular momentum conservation constrain the number of independent parameters. 

To enumerate the components of $C_{ijk\ell}$ we use the same basis for stress and strain as from Eq.~(\ref{general_stress_2d}) and Table~\ref{table_ir_2d}.
Under the assumption of isotropy, then the tensor takes the form: 
\begin{equation}
\label{eq:modfam}
\begin{gathered}
\makeatletter
\let\annualreviewGin@setfile\Gin@setfile
\let\Gin@setfile\oldGin@setfile
\includegraphics[width=7cm]{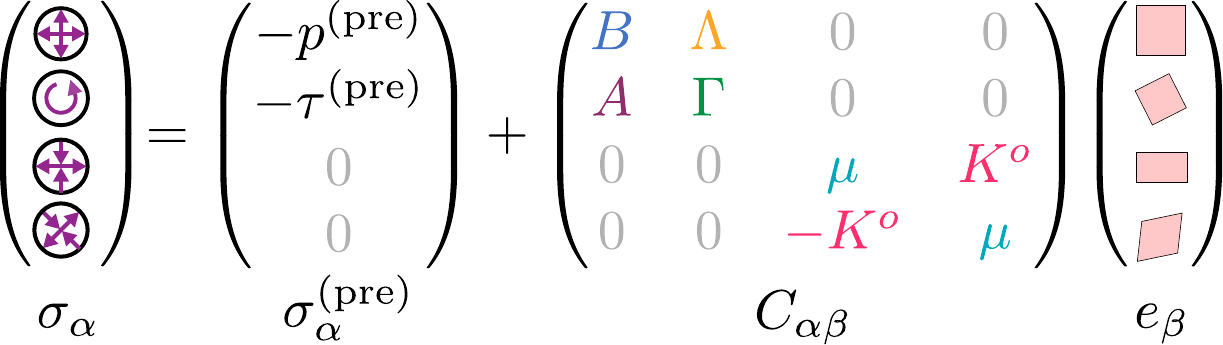}  
\let\Gin@setfile\annualreviewGin@setfile
\makeatother
\end{gathered}
\end{equation}
In Eq.~(\ref{eq:modfam}), $B$ and $\mu$ are respectively the usual bulk and shear modulus of standard isotropic elasticity. In standard tensor notation, $C_{ijk\ell}$ reads
\begin{align}
    C_{ijk\ell} =& B \delta_{ij} \delta_{k\ell}  - A \epsilon_{ij} \delta_{k\ell} - \Lambda \delta_{ij} \epsilon_{k\ell} + \Gamma \epsilon_{ij} \epsilon_{k\ell} \nonumber \\
    &+ \mu \, ( \delta_{i\ell} \delta_{jk} + \delta_{ik} \delta_{j\ell} - \delta_{ij} \delta_{k\ell})+ K^o (\epsilon_{ik} \delta_{j\ell} + \epsilon_{j\ell} \delta_{ik} )
\end{align}
Often, one also assumes that solid-body rotations do not induce stress, and hence one sets $\Lambda=\Gamma=0$. This assumption, sometimes called objectivity, is natural if the microscopic forces only depend on relative distances between points in the solid and not on the solid's orientation in space. The remaining odd elastic moduli are $A$ and $K^o$. As illustrated in Fig.~\ref{fig:elastic_casestudy}a-b, the modulus $A$ converts dilation into torque, and the modulus $K^o$ converts shear strains to shear stresses with a rotation of principal axes.

To see the relationship to energy conservation, one needs to examine the tensor $C^{\rm PK}_{ijk\ell}$, not $C_{ijk\ell}$. 
For the matrix in Eq.~(\ref{eq:modfam}), the conversion formula in \eqref{PK_C_elastic_tensors} yields
\begin{align}
C^{\rm PK}_{\alpha \beta}
= 
\mqty( B -\frac{p^\pre}2 & \Lambda+\frac{\tau^\pre}2 & 0 & 0 \\
A-\frac{\tau^\pre}2 & \Gamma-\frac{p^\pre}2 & 0 & 0 \\
0 & 0 & \mu+\frac{p^\pre}2 & K^o-\frac{\tau^\pre}2 \\
0 & 0 & -K^o+\frac{\tau^\pre}2 & \mu+\frac{p^\pre}2
) 
\end{align} 
The linear elasticity is compatible with a potential energy if and only if $C^{\rm PK}_{\alpha \beta} = C^{\rm PK}_{\beta \alpha}$. For the special case in Eq.~(\ref{eq:modfam}), this is the case if only if $A-\Lambda=2\tau^\pre=2K^o$. Fig.~\ref{fig:elastic_casestudy}c-d shows an example of a cycle of manipulations that extract energy via the moduli $A$ and $K^o$, respectively.  

\subsubsection{Elastostatics}
One immediate consequence of the presence of odd elastic moduli is that the static response of a solid changes when it is exposed to external loads and stresses. 
For example Fig.~\ref{fig:elastic_casestudy}e-f illustrates a solid under uniaxial compression~\cite{Scheibner2020}. Without odd elasticity, a typical solid will deform with left right symmetry (Panel f). However, a solid with nonzero $K^o \neq 0$ will display a chiral, horizontal deflection. The ratio of the horizontal to the vertical motion of the top surface is dubbed the odd ratio $\nu$.
Additionally, the Poisson ratio $\nu$ measures the ratio of horizontal expansion to vertical compression.
Their physical meaning is illustrated in Fig.~\ref{fig:elastic_casestudy}e-f and their values are given by 
\begin{align}
\nu^o = -\frac{\partial_y u_x}{\partial_y u_y} = \frac{B K^o}{\mu (B+\mu) + (K^o)^2} 
\quad
\text{and}
\quad
\nu = -\frac{\partial_x u_x}{\partial_y u_y} = \frac{\mu(B-\mu)-(K^o)^2}{ \mu (B+\mu)+(K^o)^2 }. \nonumber
\end{align}
For sufficiently large $K^o$, the material becomes auxetic, meaning that $\nu<0$.  

\subsubsection{Elastodynamics}
\label{sec_elastodynamics}
So far, our discussion of elasticity has not yet involved any equations of motion.
For inertial systems, the dynamics of the displacement field can be described by
\begin{align} 
\label{eom_elastodynamics}
\rho \partial_t^2 u_i + \Gamma \partial_t u_i = f_i = \partial_j \sigma_{ij} 
\end{align} 
where $\rho$ is the mass density and the term $\Gamma \partial_t u_j$ represents friction on a lubricated substrate.
Explicitly, the equations read:
\begin{align}
\rho \partial_t^2 \vec{u} + \Gamma \partial_t \vec{u} = B \nabla (\nabla \cdot \vec{u}) + \mu \Delta \vec{u} -A \epsilonb \cdot \nabla( \nabla \cdot \vec{u}) + K^o \epsilonb \cdot \Delta \vec{u} \label{eq:eom} 
\end{align} 
Since Eq.~(\ref{eq:eom}) is linear, it supports plane wave solutions  $\vec{ u}(\vec{ x}) = \vec{ u} e^{i (\vec{q} \cdot \vec{ x} - \omega t) }$ with wave number $\vec{q}$ and frequency $\omega$. It is useful to write Eq.~(\ref{eq:eom}) as a matrix equation in terms the longitudinal $u_\parallel= \hat {\vec{ q}} \cdot \vec{u}$ and transverse $u_\perp = \hat {\vec{ q}} \times \vec{ u}$ components:
\begin{align}
 -(\rho \omega^2 +i \Gamma  \omega) \mqty(u_\parallel \\ u_\perp )  = -q^2 \mqty( B+\mu  & K^o \\  -K^o+A & \mu   ) \mqty(u_\parallel \\ u_\perp )
 =  D \mqty(u_\parallel \\ u_\perp )
 \label{eq:mateom}
\end{align}
The matrix $D$ in Eq.~(\ref{eq:mateom}), known as the dynamical matrix, relates forces to displacements. 
When $\Gamma = 0$ and $A=K^o=0$, we recover the usual two types of elastic waves: a longitudinal and transverse mode, with dispersions $\omega = \pm q\sqrt{\frac{ B+\mu}{\rho}} $ and $\omega= \pm q \sqrt{\frac \mu \rho} $, respectively. The transverse wave is illustrated in Fig.~\ref{fig:elastic_casestudy}g. However, in an overdamped system ($\rho=0$ and $\Gamma >0$) passive elastodynamics ($A=K^o=0$) becomes diffusive: $\omega = -i q^2 \frac{B+\mu}\gamma$ and $\omega = - i q^2 \frac{\mu}\Gamma$. In our convention, a negative imaginary frequency implies that a wave is attenuated. When $A, K^o \neq 0$, we obtain
\begin{align}
    \omega = - i q^2 \frac{ B/2 + \mu \pm \sqrt{ (B/2)^2 - K^o (K^o-A) }  }{\Gamma} \label{eq:spectrum}
\end{align}
Notice that when $K^o (K^o-A) > (B/2)^2$, the frequency has a real part, implying oscillations even though the system is overdamped.
Just as in the case of a damped harmonic oscillator, the transition between exponential relaxation and damped oscillations is marked by an exceptional point, where the dynamical matrix $D$ is not diagonalizable~\cite{Scheibner2020} (see Refs.~\cite{Shankar2020,Ashida2020} for an introduction to exceptional points). 
In Fig.~\ref{fig:elastic_casestudy}h, an odd elastic phonon is shown for $K^o> 0$, $A=B= \mu=\rho=0$. Notice that the displacement field is circularly polarized, i.e. it traces out ellipses. A single point in the displacement field is shown as a function of time for an undamped passive solid ($A=K^o =0$) in Fig.~\ref{fig:elastic_casestudy}I, and for an overdamped odd elastic solid ($B=\mu=0$) in Fig.~\ref{fig:elastic_casestudy}J.

\begin{figure}[h]
\includegraphics[width=0.7\textwidth]{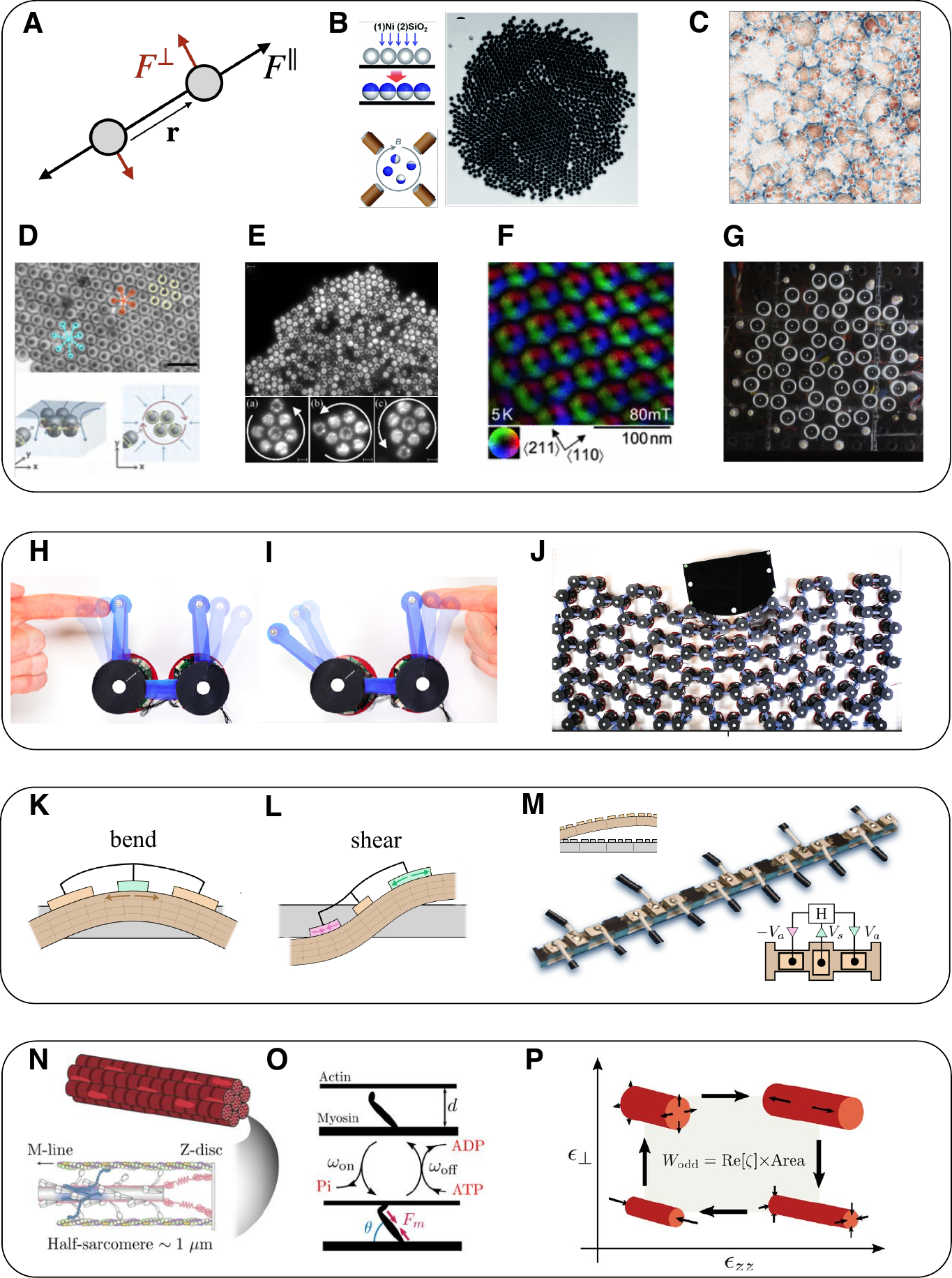}
\caption{\label{fig:elmicro}
{\bf Experimental platforms for odd elasticity.} 
({\bf A})~A pairwise interaction with a longitudinal force $F^\parallel (\vec{ r})$ and a transverse force $F^\perp (\vec{ r})$.  
({\bf B})~Janus particles and 
({\bf C}) hematite colloids driven to spin by external magnetic fields. Adapted from~\cite{Yan2015,Bililign2021}.  
({\bf D})~Starfish embryos and 
({\bf E})~bacteria~\cite{Petroff2015} form chiral crystals with particle rotation driven by flagella and celia, respectively. Adapted from~\cite{Tan2021, Petroff2015}. 
({\bf F})~Skyrmion lattices exhibit transverse interaction via a magnus force. Adapted from~\cite{Bauer2016generic}.  
({\bf G})~Gyroselastic media can be mapped onto odd elastic equations of motion in the limit of a fast spinning gyroscope. Adapted from~\cite{Nash2015} 
({\bf H-I})~A realization of the active hinge described by Eq.~(\ref{eq:hinge}).  
({\bf J})~Such hinges are tiled into a 2D wall to create an odd elastic solid. Adapted from~\cite{Brandenbourger2021Impact}. 
({\bf K-L})~A moderately thick beam with piezoelectric patches that couple bending and shearing degrees of freedom. 
({\bf M})~The repeated unit cell gives rise to a 1D chain with odd elasticity. Adapted from~\cite{Chen2021}. 
({\bf N})~A schematic depicting a muscle fiber along with 
({\bf O})~internal chemical reactions. 
({\bf P})~The active stresses can couple transverse and longitudinal strains anti-symmetrically. Adapted from~\cite{Shankar2022}. 
}
\end{figure}

\begin{figure*}[h]
\includegraphics[width=16cm]{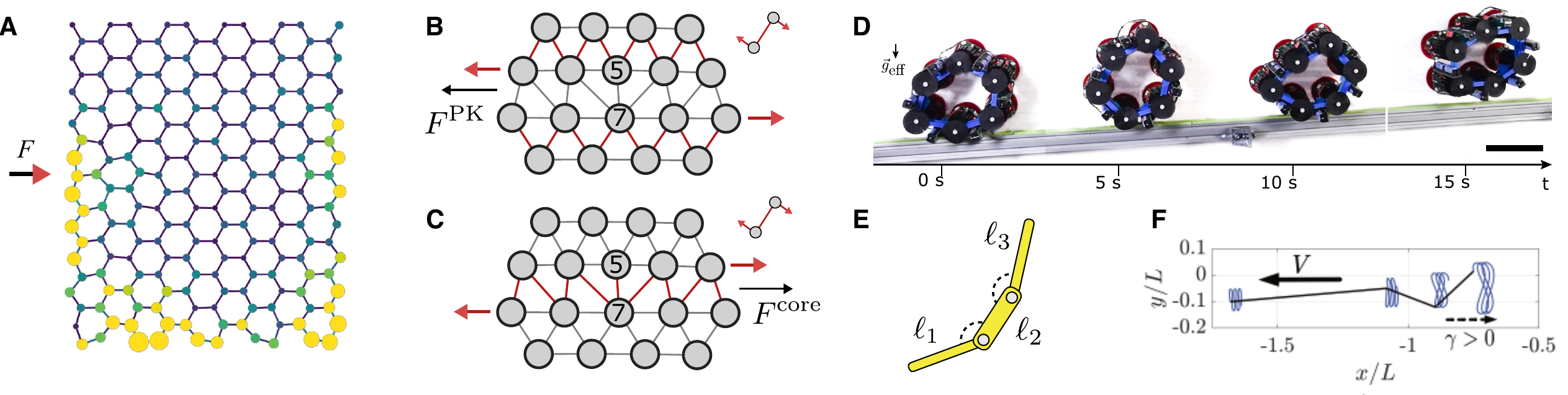}
\caption{\label{fig:elpheno}
\strong{ Phenomenology of odd elastic systems.} 
({\bf A})~A honeycomb lattice is formed from masses and springs with transverse interactions, see Eq.~(\ref{eq:inter}). When the system is poked at the side (black arrow), a unidirectional wave propagates at the boundary. Color and size are proxies for displacement. This wave is protected by a topological invariant known as the Chern number~\cite{Scheibner2020non}.
({\bf B})~A triangular lattice of bonds experiencing a clockwise transverse force forms an odd elastic solid. 
A topological defect called a dislocation is embedded in the lattice: it consists of a particle with only five neighbors (labeled 5) paired with a particle with seven neighbors (labeled 7; other particles have 6 neighbors).
These are separated by a horizontal line called the glide plane.
The transverse forces from the bonds not crossing the glide plane (highlighted in red) give rise to opposing lateral forces on the rows of atoms containing the 5 and the 7. These forces motivate the dislocation to travel left, and can be captured by a continuum notion known as the Peach-Koehler force $F^\text{PK}$.   
({\bf C})~The bonds that straddle the glide plane push the opposite direction, and therefore motivate the dislocation to move to the right. This effect has been referred to as a core force $F^\text{core}$~\cite{Braverman2021}, which evades a continuum description. The core force vanishes when the microscopic interactions are entirely longitudinal. 
({\bf D})~The active hinge from Fig.~\ref{fig:elmicro}H is configured in a hexagon. Such a hexagon undergoes an instability that propels it up a ramp. Adapted from~\cite{Brandenbourger2021Impact}.
({\bf E-F})~The same linkage system, now with different length rods, undergoes shape changes that power swimming at low Reynolds number. Time traces of the joints are shown and $V$ represents the average velocity. Adapted from~\cite{Ishimoto2022self}.} 
\end{figure*}

\subsection{Where to find odd elasticity?} \label{sec:micromodels}

We now turn to the questions: 
what experimental systems display odd elasticity? and what are the key ingredients to observe it?  
To our knowledge, no quantitative measurement of an odd elastic modulus has yet been reported in an experiment. 
Nonetheless, many systems exhibit the qualitative ingredients required for odd elasticity, and it has been suggested as an explanation for observed mechanical behavior in colloidal and biological systems~\cite{Bililign2021,Tan2021,Shankar2022}. 
Odd elastic moduli have also been determined in realistic simulations of engineered metamaterials~\cite{Brandenbourger2021Impact,Chen2021}.

\subsubsection{Point particles with pairwise interactions} \label{sec:pairwise}
The simplest microscopic picture of an elastic solid is a collection of masses connected by springs, or more generally, point particles interacting via pairwise forces~\cite{Scheibner2020,Scheibner2020non,Braverman2021,Bililign2021}. 
This class of models can be described as follows.
Let us consider $N$ particles with positions $\vec{ x}^\alpha (t)$ 
and assume that the total interaction force
$\vec{ F}^\alpha(\vec{ x}^1, \dots, \vec{ x}^N)$ 
acting on it only depends on the particle positions.
This force is said to be potential if $F_i^\alpha = - {\partial V}/{\partial x_i^\alpha}$ for some potential function $V(\vec{ x}^1, \dots, \vec{x}^N)$. Here, we will not assume, as is often done, that the forces are potential.
However, we will focus on forces that respect Newton's third law, as these can be captured by a stress tensor in the continuum (see Box~\emph{Violations of Newton's Third Law} for when this is not the case).
Formally, Newton's third law means that the forces $\vec{F}^\alpha$ can be decomposed as $\vec{F}^\alpha (\vec{x}^1 ,\dots, \vec{x}^N ) = \sum_{\beta} \vec{F}^{\alpha \beta} (\vec{x}^1, \dots, \vec{x}^N)$ with $\vec{F}^{\alpha \beta} = - \vec{F}^{\beta \alpha}$. 

For concreteness, let us focus on two dimensional systems with pairwise forces that are covariant under rotation. 
In this case, the force law takes the form $\vec{F}^{\alpha \beta} (\vec{ x}^1, \dots, \vec{x}^N) = \vec{ F} (\vec{x}^\alpha - \vec{x}^\beta)$ where 
\begin{align}
    \vec{ F} (\vec{ r})  =  F^\parallel(r) \, \hat{\vec{ r}} - F^\perp(r) \,  \bph \label{eq:inter}
\end{align}
in which $\hat {\vec{ r}} = \vec{ r}/r$, $r=\lVert \vec{r} \rVert$, and $\bph = -\epsilonb \cdot \hat {\vec{ r}}$. As illustrated in Fig.~\ref{fig:elmicro}A, $F^\parallel(r)$ is a radial force and $F^\perp(r)$ is a transverse force. 
Notice that such an interaction is compatible with a potential if and only if $\nabla \times \vec{ F} = \frac1r \partial_r (r F^\perp) =0 $. 
Except for the special case of $F^\perp \propto 1/r$ (relevant to vortices, see section \ref{skyrmions_vortices_gyroscopes}), the force is nonpotential if $F^\perp \neq 0$. 
As illustrated in Fig.~\ref{fig:elmicro}B-G, the model in Eq.~(\ref{eq:inter}) can be used to describe systems ranging from biological and colloidal crystals, to skyrmions and gyroscopic matter.

Particles interacting via Eq.~(\ref{eq:inter}) tend to form hexagonal lattices. Linearizing about a perfect lattice with spacing $a$, we may write Eq.~(\ref{eq:inter}) as
\begin{align} 
F^\parallel(r) \approx& F^\parallel(a) - k (r-a)  &  F^\perp(r) \approx& F^\perp(a) - k^a(r-a). \label{eq:linearized} 
\end{align} 
The linearized interactions can be thought of as Hookean springs with spring constant $k$ and a transverse spring constant $k^a$. When $\vec{ F}(r)$ falls off sufficiently rapidly, one can keep only the interactions between nearest neighbors. In this approximation, the ambient pressure $p^\pre$ and ambient torque $\tau^\pre$ in a hexagonal crystal are given by:
\begin{align}
    p^\pre =& \sqrt{3} \, \frac{F^\parallel (a)}{a} & 
    \tau^\pre =& -\sqrt{3} \, \frac{F^\perp(a) }{a}
\end{align}
and the isotropic 2D elastic moduli are given by:
\begin{align}
    B=& \frac{\sqrt{3}}2 \qty( k + \frac{F^\parallel(a)}{a} ) &
    \mu =& \frac{\sqrt{3}}4 \qty( k - \frac{3 F^\parallel(a)}{a} )\\
     A=& -\frac{\sqrt{3}}2 \qty( k^a + \frac{F^\perp(a)}{a} ) &
    K^o =& \frac{\sqrt{3}}4 \qty( k^a - \frac{3 F^\perp(a)}{a} ) \label{eq:modulus}
\end{align}
We see explicitly from Eq.~(\ref{eq:modulus}) that the transverse force gives rise to odd elastic moduli. 
See Refs.~\cite{Poncet2022When,Scheibner2020,Braverman2021} for more details on the coarse-graining procedures and Ref.~\cite{Chen2022} for situations in which disorder is included.

\begin{textbox}[h]\section{Gyroelasticity} \label{box:gyro}
Gyroelasticity provides a continuum description for networks of coupled gyroscopes~\cite{Wang2015,zhao2020gyroscopic,Carta2014dispersion,Hassanpour2014dynamics,Carta2017deflecting,Nash2015,Mitchell2018,Mitchell2018Realization,Mitchell2018Nature,Brun2012vortex,Benzoni2021Rayleigh}. 
Consider tracking the displacements $\vec{ x} $ of the tip of a gyroscope pinned to the ceiling. For small deflections, the tip will respond to a Newtonian force $\tilde{\vec{F}}$ as 
\begin{align}
(m \delta_{ij} \partial_t^2  + \alpha \epsilon_{ij} \partial_t  ) x_j = \tilde F_i 
\end{align}
where $\alpha$ is proportional to the angular momentum of the gyroscope~\cite{zhao2020gyroscopic}.  
In the continuum, the force is the divergence of the stress $\tilde F_i = \partial_j \tilde{\sigma}_{ij}$ where $\tilde{\sigma}_{ij} = \tilde C_{ijk\ell} \partial_\ell u_k$.
In the limit that $\alpha \to 0$, one obtains standard elasticity. In the limit that $m\to 0$, we can rewrite the equations as $ \alpha \partial_t x_i =  \epsilon_{ij} \tilde{F_j} $.
Since the motion is perpendicular to the force,  as discussed in section~\ref{skyrmions_vortices_gyroscopes},
it is useful to define an effective stress $\sigma_{ij} = \epsilon_{ik} \tilde \sigma_{kj} $. Defined this way, the ``effective" stress-strain relationship $\sigma_{ij} = C_{ijk\ell} \partial_\ell u_k$ has odd elasticity (though the stress does not correspond to linear momentum transfer). 
In general, gyroelasticity interpolates between the odd elasticity with first-order dynamics and the standard elasticity with second-order dynamics. Topological boundary modes and other chiral waves have been studied extensively in these systems~\cite{Wang2015,zhao2020gyroscopic,Carta2014dispersion,Hassanpour2014dynamics,Carta2017deflecting,Nash2015,Mitchell2018,Mitchell2018Realization,Mitchell2018Nature,Brun2012vortex,Benzoni2021Rayleigh}.
\end{textbox}

\subsubsection{Skyrmions, vortices, and gyroscopes}
\label{skyrmions_vortices_gyroscopes}
Describing the motion of particles in lossy environments (such as colloids in water) and quasiparticles such as topological defects often involves a mobility matrix  $\vec{\mu}$, such that $ \dot{\vec{ x}} = \vec{\mu} \cdot \tilde{\vec{ F}} \equiv \vec{ F}$, where $\tilde {\vec{ F}} = - \nabla V$ is potential.
Such an equation of motion arises, for example, in collections of fast-spinning, pinned gyroscopes connected by springs~\cite{Wang2015,zhao2020gyroscopic,Carta2014dispersion,Hassanpour2014dynamics,Carta2017deflecting,Nash2015,Mitchell2018,Mitchell2018Realization,Mitchell2018Nature,Brun2012vortex}, skyrmions~\cite{Benzoni2021Rayleigh,Huang2020melting,Ochoa2017Gyrotropic,Muhlbauer2009skyrmion,Yu2010Real,Brearton2021Magnetic} and vortices in superfluids~\cite{Sonin1987vortex, Gifford2008dislocation,Nguyen2020Fracton,Moroz2018effective, Fetter2009rotating,Blatter1994vortices,Tkachenko1968elasticity,tkachenko1966vortex,tkachenko1966stability}\footnote{Skyrmions and superfluid vortices are topological objects in the magnetic and velocity field, respectively. They both experience a so called Magnus force that gives rise to their transverse interactions.  
Superfluid vortices have an interaction that goes as $1/r$ for sufficiently large separation, resulting in highly complex dynamics that evade standard elasticity~\cite{Sonin1987vortex}.},  see Fig.~\ref{fig:elmicro}E-G.

If we coarse-grain the forces $\vec{ F}$, we obtain a continuum equation of motion of the form $\Gamma \partial_t  u_i = \partial_j \sigma_{ij}$ where $\sigma_{ij}= C_{ijk\ell} \partial_\ell u_k$ and $\Gamma$ is a scalar drag coefficient. If $\vec{ \mu}$ is asymmetric, then $C_{ijk\ell}$ contains odd elastic moduli. Moreover, if $\vec{\mu}$ is isotropic in two dimensions, then it is proportional to a rotation matrix $\vec{R}(\theta)$. In this case, the the nonzero odd elastic moduli are $A$ and $K^o$ and they are constrained to the ratio
$A/B = {K^o}/{\mu} = \tan \theta$.

An equivalent coarse-graining method starts with the equation of motion $[\vec{\mu}^{-1}] \cdot \dot{\vec{x}} = \tilde{\vec{F}}$ and coarse-grains the force $\tilde{\vec{ F}}$. This is a common approach for systems governed by a Lagrangian of the form $L = \sum_\alpha \vec{ x}^\alpha \cdot \epsilonb \cdot \dot{ \vec{ x}}^\alpha - V(\vec{x}^1, \dots, \vec{ x}^N)$ such as skyrmions or gyroscopes~\cite{Nguyen2020Fracton,Moroz2018effective}. We then obtain a continuum equation of motion of the form $\Gamma_{j k} \partial u_k = \partial_i \tilde \sigma_{ij}$, where $\sigma_{ij} = \tilde C_{ijk\ell} \partial_k u_\ell$ and $\vec{\Gamma} = \Gamma \vec{\mu}^{-1}$. Since $\tilde{\vec{F}}$ is conservative, the elasticity tensor $\tilde C_{ijk\ell}$ has no odd elasticity. However, the complexity has been shifted to the fact that the effective drag coefficient $\vec{\Gamma}$ is now a tensor. 
Not all combinations of odd elastic moduli can be obtained in this way. Whenever $A/B \neq K^o/\mu$, the elastic moduli are not compatible with an asymmetric mobility matrix~\cite{Scheibner2020non,Braverman2021}. 
Likewise, odd elasticity does not capture all of gyroscope mechanics: when not in the fast spinning limit, such gyroscopic systems give rise to a distinct, full-fledged theory known as gyroelasticity, see Box~\emph{Gyroelasticity}.

The systems discussed in this section share an important conceptual feature: the stress tensor $\sigma_{i j}$ does not represent the flux of linear momentum, and the dot product $\delta W =  \vec{F} \cdot  \delta {\vec{ x}}$ does not represent a physical energy.
For example, quasiparticles such as vortices or skyrmions do not have mass or linear momentum in the usual sense.  
Likewise, the motion of the tips of pinned gyroscopes is governed by the transfer of angular momentum, not linear momentum since the gyroscopes are anchored to a substrate. 
Moreover, as will be discussed in the next section, submerged particles at low Reynolds number are not ruled by linear momentum conservation, as they continually exchange momentum with the fluid.
In these situations, the equations of motion are a firmer starting point than the mechanical notion of force.

\begin{textbox}[h]\section{Violations of Newton's Third Law} \label{box:newton} 
In most of this review, we have assumed that linear momentum is conserved. 
Microscopically, this means that when two constituents interact, they exert equal and opposite forces on each other (this is Newton's third law).
At the continuum level, it means that inter-particles interactions enter the balance of linear momentum equation as the divergence of a momentum current ($\nabla \cdot \vec{\sigma}$).
However, there are many systems in which linear momentum is effectively not conserved (at the level of description that is most convenient; when all degrees of freedom are kept in the description, linear momentum of a closed system is conserved).
These include collections of self-propelled particles such as birds or active colloids~\cite{Palacci2014Light,Briand2018Spontaneously,Desreumaux2012Active,Francois2019group,Baconnier2021}, 
particles interacting through hydrodynamic interactions~\cite{Guazzelli2011Fluctuations,Beatus2007Anomalous,Beatus2006phonons,Baek2018generic,Uchida2010} or through chemical fields \cite{Soto2014,Saha2019,Meredith2020}, complex plasma \cite{Ivlev2015} and optical matter \cite{Peterson2019,
Yifat2018,
Han2018}.
For solids, the continuum force can then depend on the strain (not only on its gradients)~\cite{Poncet2022When}. 
For example, in the context of microfluidic 1D crystals, one can derive a linear wave equation of the form~\cite{Beatus2007Anomalous} 
\begin{align}
\partial_t u = \alpha \partial_x u + \beta \partial_x^2 u \label{eq:beatus}
\end{align}
The first term in Eq.~(\ref{eq:beatus}) results from an effective violation of Newton's third law and is lower order in gradients than one would expect from elasticity. In 2D crystals, violations of Newton's third law have been shown to spontaneously drive dislocation motion~\cite{Poncet2022When}. 
\end{textbox}

\subsubsection{Spinning particles at low Reynolds number: from driven colloids to starfish embryos}
The model in Eq.~(\ref{sec:pairwise}) has been used to model 2D aggregates of particles driven to spin at low Reynolds number. 
As shown in Fig.~\ref{fig:elmicro}b-e, examples include Janus particles~\cite{Yan2015}, magnetic colloids~\cite{Bililign2021}, starfish embryo~\cite{Tan2021}, and spinning bacteria~\cite{Petroff2015Fast}. 
The fluid mechanics in these systems involves a complex interplay between electromagnetic interactions, steric contact between particles or with the substrate, and particle shape change (e.g. flaggeler and cilial motion). 
Odd elasticity has been suggested~\cite{Bililign2021,Tan2021} as a natural ingredient in the continuum theory because the fluid-mediated force between two spinning particles has a nonzero transverse component~\cite{Goldman1967Brenner,happel1981low,Jager2011pattern}, thus resembling the model in Eq.~(\ref{eq:inter}). 
Open challenges include the systematic coarse-graining of the particle system with hydrodynamic interactions into an effective continuum theory, and the role of additional order parameters (such as the angular momentum of the particles) in this description.

\subsubsection{Active hinges: odd elasticity with conserved angular momentum}  
In an inertial system, the transverse force in Eq.~\eqref{eq:inter} inherently requires a torque to be provided to each bond. The building blocks must therefore have an internal or external reservoir of angular momentum. However, such a reservoir is not necessary for odd elasticity for systems that have beyond pairwise interactions~\cite{Scheibner2020}. For example, the linkage shown in Fig.~\ref{fig:elmicro}H-I has two angular degrees of freedom $\theta_1$, $\theta_2$, each determined by the location of three vertices. When an angle deforms, it experiences a torsional stiffness $\tau_i$ proportional to the change in angle $\delta \theta_i$, for example given by
\begin{align} 
\mqty( 
\tau_1 \\
\tau_2
) 
=& 
\mqty( -\kappa  &  \kappa^a  \\
-\kappa^a  & -\kappa ) 
\mqty( \delta \theta_1 \\
\delta \theta_2) \label{eq:hinge}
\end{align} 
Here $\kappa$ sets the standard bond bending stiffness provided, for example, by the stiffness of the plastic subtending the device. The coefficient $\kappa^a$ is an antisymmetric coupling. The active hinge described by Eq.~\eqref{eq:hinge} can be realized, for example, in robotic metamaterials, see Fig.~\ref{fig:elmicro}J and Ref.~\cite{Brandenbourger2021Impact}.  
For this system, the work done by the torsional stiffness is $\delta W = \tau_i \delta \theta_i$. Hence the torque-angle relationship is nonconservative when $\kappa^a\neq 0$. Consequently, a lattice made of such units (Fig.~\ref{fig:elmicro}J) will generically exhibit odd elasticity in the continuum limit.
Since the microscopic building block has no unbalanced torques, $A=0$ and $\tau^\pre=0$ in the continuum description. 

\subsubsection{Slender geometries: from beams and rings to muscles and biomembranes} \label{sec:slender} 
So far, we have primarily considered 2D isotropic media confined to the plane, but this need not be the case. 
For example, Fig.~\ref{fig:elmicro}K-M shows a quasi-1D metamaterial in which each unit cell consists of three piezoelectric patches mounted on a steel beam~\cite{Chen2021}. The beam has two modes of deformation, bending (Fig.~\ref{fig:elmicro}K) and shearing (Fig.~\ref{fig:elmicro}L). These modes of deformation in turn induce a shear stress $\sigma$ and a bending moment $M$. The constitutive relation between the two takes the form
\begin{align}
    \mqty( \sigma \\ M ) = \underbrace{ \mqty( \mu & P \\ 0 & B)}_{C} \mqty( s \\ b) \label{eq:const}
\end{align}
The matrix $C$ plays the role of the elastic modulus tensor, and $\mu$ and $B$ are the shear and bending moduli, respectively, that one would expect from Timoshenko-Ehrenfest beam theory~\cite{Timoshenko1940strength}. 
An electronic feedback between the piezoelectrics induces an additional modulus $P$. Since the energy differential is $\delta W  = \sigma \delta s + M \delta b $, the asymmetric part of $C$ corresponds to a violation of Maxwell-Betti reciprocity and therefore requires a source of energy.  
Since $P$ violates parity and is nonconservative, it also induces unidirectional wave amplification. 

Theoretical designs for 2D piezoelectric odd elastic materials have been proposed~\cite{Cheng2021}. 
In addition to quasi-1D structures~\cite{Brandenbourger2021Impact,Chen2021}, three dimensional solids~\cite{Scheibner2020,Shankar2022}, thin membranes in~\cite{Salbreux2017} and moderately thick plates have been considered.
Odd elasticity can also emerge, in principle, from more complex building blocks. For example, as illustrated in Fig.~\ref{fig:elmicro}N-P, it has been recently suggested that muscle tissue in suitable operating regimes can be approximately modeled as a 3D anisotropic odd elastic medium~\cite{Shankar2022,Zahalak1996}.

\subsection{Odd elastic phenomenology}

\subsubsection{Topological waves} 
Nonconservative forces can be useful for constructing mechanical systems with nontrival band topology (see \cite{Shankar2020,Ashida2020,Fruchart2013} for introductions to topological waves). 
Networks of bonds obeying Eq.~(\ref{eq:inter}) can behave as Chern insulators~\cite{Scheibner2020non,Zhou2020non, Scheibner2020}, where unidirectional edge modes can propagate at the boundary of the system (see Fig.~\ref{fig:elpheno}A).
Since the forces are nonconservative, the dynamical matrix governing the linear waves is in general non-Hermitian, leading to features such as the non-Hermitian skin effect, in which the bulk modes are localized at one edge of a 1D system, and exceptional points, in which the eigenvectors of the dynamical matrix do not span the Hilbert space~\cite{Ashida2020,Shankar2020}.  
For instance, unidirectionally amplified waves arising from the non-Hermitian skin effect have been observed in the beam in Fig.~\ref{fig:elmicro}K~\cite{Chen2021} and in 1D chains of the active hinges in Fig.~\ref{fig:elmicro}H~\cite{Brandenbourger2021Impact}. 
We refer to \cite{Shankar2022} and references therein for more details.

\subsubsection{Topological defects}
A crystalline topological defect is an imperfection in a crystal structure that cannot be removed by local rearrangements of the particles. 
These defects act as quasi-particles, and their motion governs the large scale rearrangement of the crystal~\cite{Nelson2002, Weertman1964,Nelson1979}. 
For example, in a hexagonal lattice, an extra row of atoms inserted into the crystal structure typically results in a pair of atoms with 5 and 7 Voronoi neighbors, known as a dislocation, see~Fig.~\ref{fig:elpheno}B. 
The dislocation carries a topological charge, the Burgers vector $b_j = \oint \partial_i u_j \dd r_i $, where the contour encloses the dislocation of interest. 

The presence of transverse forces as in Eq.~(\ref{eq:inter}) fundamentally modifies how the defects move and interact. 
A dislocation is said to glide when the two rows of atoms containing the 5 and 7 fold coordinated particle slide past each other. The dislocation motion itself is the direction of motion of the row containing the 5 fold particle. 
For forces with a clockwise handedness, the bonds that do not cross the glide plane favor dislocation motion to the right, as shown in Fig.~\ref{fig:elpheno}B.  However, the bonds that cross the glide plane push the dislocation in the opposite direction, see Fig.~\ref{fig:elpheno}C. 
These two effects are in competition.
The effect of bonds that do not cross the glide plane can be captured by a continuum notion known as the Peach-Koehler force $F^{\rm PK}_i =  -b_k\sigma_{kj}^\pre \epsilon_{ji} $, where $b_k$ is the Burgers vector and $\sigma^\pre_{ij} = \tau^\pre \epsilon_{ij}$ is the torque density induced by microscopic transverse forces. The contribution from the bonds that cross the glide plane evades a continuum explanation, because it arises from a difference in forces separated by one lattice spacing $a$.
It can be attributed to a core force $F^\text{core}$ that can be computed from microscopic data as $F^\text{core} = \frac1a\int_{\mathcal{C}} \vec{F}(\vec{r}) \cdot \dd \vec{r}$, where $\vec{ F}$ is the microscopic interaction mediated by the bonds and $\mathcal{C}$ is the concatenation of the trajectories of all the bonds that cross the glide plane when the dislocation moves by a single unit cell~\cite{Braverman2021}.
Transverse forces have been shown in simulation and experiment to cause spontaneous dislocation glide and nucleation~\cite{Bililign2021,Scheibner2020}.  
The interactions between topological defects (mediated by elastic strains) does not fall within the typical conservative field theory paradigms. 
An open question, relevant for systems ranging from skyrmion lattices~\cite{Huang2020melting} to complex fluids~\cite{Bililign2021,Yan2015}, is how modified defects dynamics affect the nature of plastic deformation and melting.

\subsubsection{Nonlinearities and noise}
In solids with energy injection, instabilities are generic. For example when $\rho\neq0$ and $K^o$ is sufficiently large, the linear odd elastic waves described by  Eq.~(\ref{eq:mateom}) become unstable~\cite{Scheibner2020}. 
A linear instability either results is destruction of solid material or it must be stabilized by nonlinearities. 
Nonlinear dynamical structures such as limit cycles can prove useful. 
For example, in Ref.~\cite{Brandenbourger2021Impact}, the active hinge when put on a ring can locomote up an incline (see Fig.~\ref{fig:elpheno}D-F). 
In Refs.~\cite{Ishimoto2022self,Yasuda2021Odd} the active hinge is immersed in a viscous fluid, and the limit cycle results in swimming behavior (see Fig.~\ref{fig:elpheno}G-H). 
The interplay between nonconservative forces and noise has also been studied~\cite{Yasuda2022a,Yasuda2022b}. 
In the context of the swimming hinges, it is shown that random fluctuations give rise to (on average) persistent motion~\cite{Yasuda2021Odd,Ishimoto2022self}.

\subsubsection{Odd viscoelasticity} 
In general, a material's linear response need not be instantaneous in time. 
This can be captured by the equation
\begin{align}
\sigma_{ij} (t) = \int_{-\infty}^{\infty} M_{ijk\ell}(t-\tau) \partial_\ell \dot u_k (\tau) \dd \tau  
\end{align}
in which a finite-time response kernel $M_{ijk\ell}(t)$ has been introduced, in a way similar to electrodynamics in materials~\cite{landau2013electrodynamics}. 
Viscosity and elasticity are special cases with $M_{ijk\ell}(\tau) = \eta_{ijk\ell} \delta(\tau)  $ and $M_{ijk\ell} (\tau)  = C_{ijk\ell}$, respectively. 
The work done over a closed cycle deformation of period $T$ is given by~\footnote{For simplicity, we are ignoring the distinction between the Piola-Kirchhoff and the Cauchy stress in this discussion.}
\begin{align}
    W = \int_0^T \sigma_{ij} \partial_j \dot u_i (t) \dd t = - i  \sum_\omega  \omega  \sigma_{ij}(\omega) \partial_j \bar u_i (\omega) = - \sum_{\omega} \omega^2 \partial_j  \bar u_i (\omega)  M_{ijk\ell} (\omega) \partial_\ell  u_k (\omega) 
\end{align}
where the sum is over $\omega = 2\pi n/T$ for integer $n$~\cite{Bililign2021,Shankar2022,Banerjee2021, Chen2021}. When the material is passive, then the work must be non-positive for all possible cycles, which implies that $M_{ijk\ell}(\omega) $ (viewed as a linear operator on rank two tensors) must be positive semi-definite for all frequencies $\omega$~\cite{Muhlestein2016Reciprocity,Day1971Time,Srivastava2015Causality}. In the standard terminology of rheology, $M_{ijk\ell}(\omega)$ is often written as $i \omega M_{ijk\ell} (\omega)  = G'_{ijk\ell}(\omega) + i G''_{ijk\ell}(\omega)$, where $G'_{ijk\ell}$ is the storage modulus, containing the elasticity, and $G''_{ijk\ell}$ is the loss modulus, containing the viscosity. The adjectives ``loss" and ``storage" are misnomers when antisymmetric terms are present: the antisymmetric part of $G''_{ijk\ell}$ does not contribute to dissipation, and the antisymmetric part of $G'_{ijk\ell}$ is capable of injecting or dissipating energy. Importantly, at finite frequency, passive materials can still have an antisymmetric component of $G'_{ijk\ell}$ so long as $M_{ijk\ell}$ remains positive definite.  
Odd viscoelasticity has been considered in the context of minimal spring-dashpot models~\cite{lier2021passive}, instabilities in chiral crystals~\cite{Bililign2021}, quantum Hall effects with a tilted magnetic field~\cite{Offertaler2019}, and in simplified models of muscles~\cite{Shankar2022,Zahalak1996}. The competition between viscosity and elasticity has been suggested as a mechanism for wavelength selection at the onset of instability~\cite{Scheibner2020,Bililign2021}. Notably, $G'_{ijk\ell}$ and $G''_{ijk\ell}$ have been measured in numerical simulations of chiral active fluids~\cite{Han2021}.   

\section{Conclusion}
In this review, we explored fluid and solid mechanics in which viscosity does not dissipate energy and elasticity does not store it. 
Realistic models of complex media will almost always involve additional affects, e.g. antisymmetric stress, other active stresses, coupled fields, and strong nonlinearities. 
From the point of view of phenomenological modeling, the virtues of the odd viscosity and odd elasticity formulations lie in their simplicity: They only require shape change (i.e. diffeomorphisms) as a degree of freedom and momentum conservation. 
By combining advances in statistical mechanics, hydrodynamics, dynamical systems, and experimental methods, odd viscosity and odd elasticity may provide a window into the universal behavior of matter at long length scales.  

\section*{DISCLOSURE STATEMENT}
The authors are not aware of any affiliations, memberships, funding, or financial holdings that might be perceived as affecting the objectivity of this review. 

\section*{ACKNOWLEDGMENTS}
M.F. acknowledges support from a MRSEC-funded Kadanoff–Rice fellowship (DMR-2011854) and the Simons Foundation.
C.~S.~acknowledges support from the Bloomenthal Fellowship and the National Science Foundation Graduate Research Fellowship under Grant No.~1746045. V.V.~acknowledges support from the Simons Foundation, the Complex Dynamics and Systems Program of the Army Research Office under grant W911NF-19-1-0268, and the University of Chicago Materials Research Science and Engineering Center, which is funded by the National Science Foundation under Award No.~DMR-2011854.

%



\end{document}